\documentstyle[11pt,epsfig]{article}

\def\laq{\raise 0.4ex\hbox{$<$}\kern -0.8em\lower 0.62
ex\hbox{$\sim$}}
\def\gaq{\raise 0.4ex\hbox{$>$}\kern -0.7em\lower 0.62
ex\hbox{$\sim$}}

\begin{document}

\begin{titlepage}
\begin{flushright}
CERN-PH-TH/2005-043
\end{flushright}

\vspace{0.8cm}

\begin{center}

\huge{Cosmological perturbations for imperfect fluids}

\vspace{0.8cm}

\large{Massimo Giovannini \footnote{e-mail address: massimo.giovannini@cern.ch}}

\normalsize

\normalsize
\vspace{0.3cm}
{{\sl Centro ``Enrico Fermi", Compendio del Viminale, Via 
Panisperna 89/A, 00184 Rome, Italy}}\\
\vspace{0.3cm}
{{\sl Department of Physics, Theory Division, CERN, 1211 Geneva 23, Switzerland}}
\vspace*{2cm}

\begin{abstract}
\noindent
Interacting fluids, endowed with bulk viscous stresses,
are discussed in a unified perspective with the aim 
of generalizing the treatment of cosmological perturbation 
theory to the case where both fluctuating decay rates and fluctuating 
bulk viscosity coefficients are simultaneously present in the relativistic plasma.
A gauge-invariant treatment of the qualitatively new phenomena 
arising in this context is provided.  
In a complementary approach, faithful  gauge-fixed 
descriptions of the gravitational and hydrodynamical fluctuations are developed and exploited. To deepen the interplay between bulk viscous stresses and fluctuating decay rates, illustrative examples are proposed and 
 discussed both analytically and numerically. 
Particular attention is paid to the coupled evolution of curvature and entropy fluctuations when, in the relativistic plasma,  at least one of the interacting fluids  possesses a fluctuating bulk viscosity coefficient.  
It is argued that this class of models may be usefully employed as an effective description of the decay of the inflaton as well as of other phenomena involving imperfect relativistic fluids.
\end{abstract}
\end{center}

\end{titlepage}
\newpage
\renewcommand{\theequation}{1.\arabic{equation}}
\section{Formulation of the problem}
\setcounter{equation}{0}
It is plausible, as observations suggest, that the large-scale 
temperature fluctuations detected in the microwave sky 
may arise, via the Sachs--Wolfe effect, from primordial 
curvature  perturbations amplified during the early stages of the 
life of the Universe. It is therefore mandatory for those 
interested in this possibility to scrutinize 
in detail the evolution of curvature fluctuations when the 
composition of the primeval plasma cannot be parametrized 
in terms of a single, perfect, relativistic fluid. 

A class of dissipative effects that can enter the energy-momentum 
tensor without spoiling the isotropy of the background geometry 
can be  parametrized in terms of a bulk viscosity coefficient \cite{weinberg}
(see also \cite{LLF}). In fact, in the relativistic theory of imperfect fluids, the 
dissipative effects can be parametrized in terms of a shear coefficient, a heat 
flow coefficient and a bulk viscosity coefficient. When 
written in general coordinates \cite{weinberg} (see also \cite{bel1}), the contributions of the shear tensor and of the 
heat flow (unlike  the bulk viscosity) do not contribute to  a homogeneous and isotropic background geometry of Friedmann-Robertson-Walker type \cite{weinberg}.

Following then Ref. \cite{weinberg}, for a {\em single} fluid, the total energy-momentum 
${\cal T}_{\mu}^{\nu}$ tensor can then be split
into a perfect contribution, denoted in the following by $T_{\mu}^{\nu}$,
and into an imperfect contribution, denoted by $\Delta T_{\mu}^{\nu}$, i.e. 
\begin{equation}
{\cal T} _{\mu}^{\nu} = T_{\mu}^{\nu} + \Delta T_{\mu}^{\nu},
\label{BV1}
\end{equation}
In general coordinates, and within our set of conventions, the 
contribution of bulk viscous stresses can be written, in turn, as 
\begin{equation}
\Delta T_{\mu}^{\nu} = \xi \biggl( \delta_{\mu}^{\nu} - u_{\mu}u^{\nu}\biggr) \nabla_{\alpha} u^{\alpha},
\label{BV2}
\end{equation}
where $\xi$ represents the bulk viscosity coefficient. 
The presence of bulk viscosity 
can also be interpreted, at the level of the background, 
 as an effective redefinition of the enthalpy 
(i.e. $(\rho + p)$) that appears both in the spatial components 
of Einstein equations and in the background form of the covariant 
conservation equation.  While this approach may be 
appropriate at the level of the evolution of the background space-time, it
could be confusing if applied to the dynamics of the fluctuations 
when $\xi$ is allowed to have a spatial dependence.

The presence of bulk viscosity in the relativistic plasma 
 can influence the character of the cosmological 
singularity \cite{bel1,murphy1} (see also \cite{bel2,murphy2}).
Regular solutions \cite{murphy1} 
(but generically unstable \cite{bel1} and probably not past-geodesically 
complete) can be found if the $\xi$ has some particular 
functional dependence upon the energy density. 
Some of these solutions have the property that
as $t\to -\infty$ ($t$ being the cosmic time coordinate) 
the scale factor approaches a quasi-de Sitter stage of 
accelerated expansion. These deflationary solutions \cite{barrow1}
violate, initially, the strong energy condition, i.e. 
$(\rho + 3 {\cal P}) \geq 0$, where $ {\cal P} = p - 3\,H \xi$ is the 
total pressure and $p$ is the pressure arising from the perfect 
contribution (i.e. $T_{i}^{j} = - p \delta_{i}^{j}$) introduced 
in Eq. (\ref{BV1}). 
The dominant energy condition \cite{hawking} 
could be violated when 
bulk viscous stresses are added \cite{barrow1}. This
second occurrence implies that the effective enthalpy, i.e. $(\rho + {\cal P})$, 
may not always be positive-definite in all the classes of solutions 
involving a homogeneous (viscous) background.
Boundary conditions can be imposed in such a way 
that for $t\to +\infty$ the solution approaches, asymptotically, 
de Sitter space-time. In the latter case the solutions are driven 
to an inflationary phase \cite{barrow1}.

Barrow \cite{barrow1} presented detailed studies 
(see also \cite{barrow2,barrow3}) on inflationary Universes 
driven by a gas of strings showing that, in some regimes, this system 
can be dicussed in terms of an effective bulk viscosity coefficient.
The solutions quoted in the previous paragraphs 
have been generalized along different lines. In \cite{barrow1}
the bulk viscosity coefficient $\xi$ is taken to be proportional 
to a power of the energy density, i.e. $ \xi \propto \rho^{m}$. 
 The relevance of these examples for the cosmic no-hair 
theorems has also been investigated in a series of papers 
\cite{barrow2,barrow3}.  Moreover it was argued that, as one traces 
the Universe back in time, a point is reached when the 
expansion rate is so fast that the rate of string creation due to quantum effects 
balances the dilution of the string density due to the expansion \cite{turok1}.
These considerations lead naturally to an effective bulk viscosity 
coefficient that may induce an approximate, weak, inflationary dynamics.

The possibility of driving a robust inflationary phase by means 
of bulk viscous stresses has been indeed criticized on various grounds
(see \cite{pacher,maartens2,zim2}). In \cite{pacher} it has been argued 
that the bulk viscosity effects associated with a gas of weakly 
interacting particles cannot make the pressure negative excluding 
any form of inflation. In \cite{maartens2} it has been suggested that 
hydrodynamic inflation is intrinsically problematic (see also \cite{zim2} 
for an interesting, related perspective) since 
while the inflationary evolution requires a very small expansion 
time the hydrodynamic regime requires a very small interaction 
time.  Therefore, to treat consistently the problem one would need 
a consistent (non-linear) generalization of the Israel-Stewart transport 
equations (see the attempt of Ref. \cite{mendez} ) 
as well as a consistent model of fluid behaviour
under super-rapid expansion and strongly non-linear conditions.
The possibility of realizing a sufficiently long and robust 
inflationary phase is, however, not central for the present analysis: in the present context the possible effects of bulk viscosity will arise 
after the end of a (conventional or unconventional) inflationary 
phase.

The effect of bulk viscous stresses has also been studied  
in the case of scalar-tensor theories of gravity (see for instance \cite{cataldo}).
More recently, the possible implications of bulk viscosity on the dynamics 
of the dark energy fluid have been investigated in different 
frameworks \cite{de1} (see also \cite{de2} 
and references therein). In \cite{de1} the hope is to obtain naturally an accelerated 
expansion from bulk viscous effects.

An apparently unrelated effect  
is the energy exchange between different  fluids whose mixture constitutes the 
global relativistic plasma that drives the evolution of the geometry. 
 For instance, it is conceivable that, after inflation, the oscillating inflaton 
 field decays into massless particles \cite{KS}. Then this situation 
 can be modelled by a fluid of dusty matter decaying into a 
 fluid of radiation.  Multifluids also arise
 is the discussion of isocurvature modes like, for instance, the CDM-radiation 
 isocurvature mode. In all these problems, one of the relevant aspects 
 is the analysis of the evolution of curvature and entropy perturbations.
 
 The theory of cosmological perturbations in the presence 
 of different fluids has a long history. For instance, in \cite{KS} 
 the problem of multifluids was analysed in the absence
 of bulk viscosity.  In  \cite{mark}
multifluid systems have been 
treated in the framework of a covariant cosmological perturbation 
 (see \cite{tsagas} for an introduction and \cite{chall} 
for related applications).  Entropy perturbations 
arising in multifluid systems have also been  
discussed in \cite{hwang, nambu,malikwands,wandsmalik} in the context of the Bardeen formalism \cite{bardeen}, as well as  in the context 
of inflationary models \cite{vantent,ferr} and quintessence models \cite{noh,liddle,gordonhu,liu}. Possible effects 
connected with the decay of the inflaton have recently been considered in \cite{hector1,hector2}. Notice finally that earlier works in the general 
area of bulk viscosity can be found in \cite{maa1,maa2}.

It has been argued, in recent years, that interacting (multiple) fluids may play 
some relevant r\^ole in cosmological perturbation theory. 
If the early Universe contains more than a single fluid, the spatial 
variations of the decay rate may indeed generate curvature perturbations
 \cite{dvali} (see also \cite{kofman} and 
\cite{tsu,postma,mazumdar,averdi}).

In the present paper a unified description of imperfect fluids 
is developed with the aim of understanding if and how 
the effect of spatial fluctuations in the bulk viscosity coefficient
 (combined with a non-vanishing decay rate) may induce super-horizon 
curvature perturbations. In this sense, we are not interested in
discussing the evolution of the fluctuations in the case when a 
{\em single} viscous fluid is present.
Let us phrase the problem in physical terms. Consider 
the situation where the matter content of the Universe is formed by, at least,
{\em two fluids}. The possible dissipative effects 
compatible with the isotropy of the background geometry are:
\begin{itemize}
\item{} energy exchange between the fluids;
\item{} bulk viscous effects associated with each fluid 
and with the mixture.
\end{itemize}

The first type of efffect can be  parametrized by a (time-dependent)
decay rate that is allowed to fluctuate spatially.  These fluctuations 
 may then induce fluctuations on super-horizon scales (see, again, 
 \cite{dvali,kofman,tsu,postma,mazumdar,averdi}).
The second type of effects can be parametrized by 
the appropriate second viscosity coefficients of each fluid 
of the mixture and by their related fluctuations. Also in this 
case it is reasonable to expect that the fluctuations 
of the second viscosity coefficient can be converted 
into super-horizon curvature fluctuations.  There may be, in principle,
two extreme situations. In one case the presence 
of bulk viscous stresses crucially determines 
the nature of the background geometry. This is the approach 
studied, for instance, in \cite{murphy1,barrow1, turok1,cataldo}.
In the other situation, the bulk viscosity does not 
determine crucially the properties of the background 
geometry. This approach is hard to justify for a 
single (non decaying) viscous fluid, but it becomes 
plausible when there are at least two interacting fluids. 
Consider, for instance, the situation where the fluid 
describing the decay products has no intrinsic viscosity, while 
the fluid that is effectively decaying has some intrinsic 
viscosity. If, as in \cite{barrow1}, $\xi$
depends on the energy density of the decaying fluid, it reasonable 
to expect that when the decay is completed, the bulk viscous 
effects will also decay.  Depending on the specific 
parameters of the example, it is then conceivable that, the net result
of the evolution will be a single inviscid fluid supplemented 
by an exponentially decaying (viscous) component with 
negligible effect on the dynamics of the background. 
It is interesting to understand  what 
happens to the dynamics of the fluctuation. If the decaying 
fluid has some intrinsic viscosity, the effect on the curvature 
fluctuations may be relevant even if, when the decay is completed, 
the viscous contribution goes exponentially to zero at a rate 
that depends upon the relative interaction of the fluids of the mixture.

From a more formal point of view, one of the problems is
 a full gauge-invariant treatment of imperfect cosmological 
perturbations where  the fluctuations of  both the bulk viscosity 
coefficient and  the decay rate are invariant under 
infinitesimal diffeomorphisms. The gauge-invariant 
approach will be complemented with gauge-fixed 
descriptions that are particularly useful in connection 
with the intricacy of the numerical calculations.

The present article is organized as follows. In Section 2 the basic 
 framework of the analysis will be illustrated. Section 3 contains 
 the gauge-invariant description of cosmological perturbations 
 in the presence both of a fluctuating decay rate 
 and of a fluctuating bulk viscosity coefficient.  In Section 4 the 
 gauge-invariant evolution equations for the curvature perturbations 
 will be introduced, while in Section 5 curvature perturbations 
 will be discussed in the context of a gauge-fixed description.
 Section 6 contains various numerical examples 
 clarifying the interplay between fluctuating bulk viscosity and fluctuating 
 decay rate. Finally,  Section 7 contains the summary of the main findings 
 and the concluding discussions.
For completeness, the analysis of the evolution of the 
vector modes has been included  in an appendix. 

\renewcommand{\theequation}{2.\arabic{equation}}
\section{Basic considerations}
\setcounter{equation}{0}
Some physical quantities relevant to the problem at hand will now be introduced. 
As already pointed out, there are two independent physical effects 
that may compete in a mixture of relativistic fluids: 
the presence of a (fluctuating) decay rate and the 
presence of a (fluctuating) bulk viscosity coefficient. 

Consider a relativistic plasma formed by, at least, two species.
The associated energy-momentum tensors 
can be written as \footnote{The signature of the four-dimensional metric adopted here 
is mostly minus, i.e. $(+,-,-,-)$. Latin subscripts are used to distinguish  
the pressures and energy densities of different fluids; tensor  indices are instead 
denoted by lowercase Greek letters.}
\begin{eqnarray}
&&T_{{\rm a}}^{\mu\nu} = ( p_{\rm a} + \rho_{\rm a}) u_{\rm a}^{\mu} u_{\rm a}^{\nu} - p_{\rm a} g^{\mu\nu},
\nonumber\\
&& T_{{\rm b}}^{\mu\nu} = ( p_{\rm b} + \rho_{\rm b}) u_{\rm b}^{\mu} u_{\rm b}^{\nu} - p_{\rm b} g^{\mu\nu}.
\label{tmnab}
\end{eqnarray}
If the fluids are decaying one into one another (for instance the a-fluid  decays into the b-fluid),  the covariant conservation equation only applies to the global relativistic plasma, while the energy-momentum tensors of the single species 
are not covariantly conserved and their specific form accounts for the 
transfer of energy between the a-fluid and the b-fluid:
\begin{eqnarray}
&& \nabla_{\mu} T^{\mu\nu}_{\rm a} = - \Gamma g^{\nu\alpha} u_{\alpha} ( p_{\rm a} 
+ \rho_{\rm a}),
\nonumber\\
&&\nabla_{\mu} T^{\mu\nu}_{\rm b} = \Gamma g^{\nu\alpha} u_{\alpha} ( p_{\rm a} 
+ \rho_{\rm a}),
\label{atob}
\end{eqnarray}
where the term $\Gamma$ is the decay rate that can be both space- and 
time-dependent; in Eqs. (\ref{atob}) 
$u_{\alpha}$ represents the (total) peculiar velocity field.
Owing to the form of Eqs. (\ref{atob}), it is clear that the total energy-momentum 
tensor of the two fluids, i.e. $T_{\rm tot}^{\mu\nu} = T^{\mu\nu}_{\rm a} + T^{\mu\nu}_{\rm b}$ is indeed covariantly conserved.

Equations (\ref{atob}) can be easily generalized to the description
of more complicated dynamical frameworks, where the relativistic 
mixture is characterized by more than two fluids. 
Consider the situation where the a-fluid 
decays as ${\rm a} \to {\rm b} + {\rm c}$. Then, if a fraction $f$ of the a-fluid 
decays into the b-fluid and a fraction $(1-f)$  into the c-fluid, 
Eqs. (\ref{atob}) can be generalized as 
\begin{eqnarray}
&& \nabla_{\mu} T^{\mu\nu}_{\rm a} = - \Gamma \,g^{\nu\alpha} \,u_{\alpha} ( p_{\rm a} 
+ \rho_{\rm a}),
\nonumber\\
&& \nabla_{\mu} T^{\mu\nu}_{\rm b} = f \Gamma \, g^{\nu\alpha}\, u_{\alpha} ( p_{\rm a} 
+ \rho_{\rm a}),
\nonumber\\
&&\nabla_{\mu} T^{\mu\nu}_{\rm c} = (1 -f) \Gamma\, g^{\nu\alpha} \, u_{\alpha} ( p_{\rm a} + \rho_{\rm a}),
\label{atobc}
\end{eqnarray}
and so on.  

It is appropriate to recall here that the authors of Ref. \cite{KS} choose 
to parametrize the coupling between the fluids of the mixture through 
a more generic function. For instance, in the notations of \cite{KS}, 
the source terms of Eqs. (\ref{atob}) will simply be $Q^{\nu}_{\rm a}$ and 
$Q^{\nu}_{\rm b}$ bound by the condition\footnote{See, for instance, 
Eqs. (5.1), (5.2) and (5.3) of \cite{KS}, page 28.} $Q^{\nu}_{\rm a} + Q^{\nu}_{\rm b}=0$. The latter condition follows from the covariant conservation 
of the total energy-momentum tensor of the mixture and it is also satisfied 
by the source terms in Eq. (\ref{atob}).
The second condition 
satisfied by the source terms is that their spatial component has to be 
zero when evaluated on the Friedmann-Robertson-Walker background.
In the present approach, the spatial part of the source 
terms given in Eq. (\ref{atob}) automatically vanish on the Friedmann-Robertson-Walker background since
 $g^{\mu\nu} \overline{u}_{\mu} \overline{u}_{\nu} =1$ (implying $\overline{u}_{0}=1$ and $\overline{u}_{i} =0$).

The second relevant dissipative effect that can enter the energy-momentum 
tensor without spoiling the isotropy of the background geometry, as 
discussed in the previous section, is the one parametrized 
in terms of bulk viscosity. Following the notation of Eqs. (\ref{BV1}) 
and (\ref{BV2}),  the effective energy-momentum tensors of the two 
fluids can be written as 
\begin{eqnarray}
&& {\cal T}^{\mu\nu}_{\rm a} = T^{\mu\nu}_{\rm a} + \Delta T_{\rm a}^{\mu\nu},
\nonumber\\
&& {\cal T}^{\mu\nu}_{\rm b} = T^{\mu\nu}_{\rm b} + \Delta T_{\rm b}^{\mu\nu}.
\label{sum}
\end{eqnarray}
Hence, the Einstein equations can be written as 
\begin{equation}
R_{\mu}^{\nu} - \frac{1}{2} \delta_{\mu}^{\nu} R = 8\pi G {\cal T}_{\mu}^{\nu},
\label{EIN1}
\end{equation}
where, 
\begin{equation}
{\cal T}^{\mu\nu} ={\cal T}^{\mu\nu}_{{\rm a}} + {\cal T}^{\mu\nu}_{{\rm b}}.
\label{TOTV}
\end{equation}
According to Eqs. (\ref{BV1}),(\ref{BV2}) and (\ref{tmnab}),
 ${\cal T}_{\rm a}^{\mu\nu}$ and ${\cal T}_{\rm b}^{\mu\nu}$ will have an inviscid 
background value (characterized by $p_{\rm a, b}$ and $\rho_{\rm a, b}$) 
and a viscous correction parametrized, at the level of the background, 
by $\overline{\xi}_{\rm a}$ and $\overline{\xi}_{\rm b}$.

In the following the background geometry will be taken to be of Friedmann--Robertson--Walker  (FRW) type.
Moreover, it will be assumed that the background geometry is spatially flat so that 
the line element can be written as 
\begin{equation}
ds^2 = dt^2 - a^2(t) d\vec{x}^2,
\label{lineel}
\end{equation}
where $t$ is the cosmic time coordinate. Therefore,
 Eq. (\ref{EIN1}) implies\footnote{We denote by an overdot 
a derivation with respect to the cosmic time-coordinate $t$.} 
\begin{eqnarray}
&& H^2 = \frac{8\pi G}{3} \rho,
\label{B1}\\
&& \dot{H} = - 4\pi G ( \rho + {\cal P}),
\label{B2}
\end{eqnarray}
where $H= \dot{a} /a$ is the Hubble rate; $\rho= \rho_{\rm a} + \rho_{\rm b}$ and 
$p= p_{\rm a} + p_{\rm b}$ denote the total (inviscid) energy and pressure 
densities while ${\cal P}$ is the effective total pressure 
containing the viscous contribution, i.e. 
\begin{equation}
{\cal P} = p - 3 \overline{\xi}\, H\equiv (p_{\rm a} + p_{\rm b}) - 3 (\overline{\xi}_{\rm a} + 
\overline{\xi}_{\rm b} )\, H.
\label{effP}
\end{equation}
In the following sections, the total background viscosity coefficient will be denoted 
by $\overline{\xi}$, i.e. 
\begin{equation}
\overline{\xi} = \overline{\xi}_{\rm a} + \overline{\xi}_{\rm b} + ...,
\end{equation}
where the ellipses stand for the possible contribution of other fluids.
The definitions of the effective pressures for each fluid follows from Eq. (\ref{effP})
and it is 
\begin{equation}
 {\cal P}_{\rm a} = p_{\rm a} - 3 \overline{\xi}_{\rm a}\,H, \qquad
{\cal P}_{\rm b} = p_{\rm b} - 3 \overline{\xi}_{\rm b}\,H .
\label{effP2}
\end{equation}
Notice, furthermore,
that $\overline{\xi}$ (and, similarly $\overline{\xi}_{\rm a}$ and $\overline{\xi}_{\rm b}$)
denotes the background values of the bulk viscosity coefficients. In different physical 
examples (see e.g. \cite{murphy1,barrow1} and also \cite{turok1})
 the bulk viscosity coefficient is proportional to a power of the energy density.
If the a- and b-fluids would not be allowed to exchange energy and momentum,
the mixture could be described by a single total fluid with effective pressure ${\cal P}$.
If the decay rate does not vanish, Eqs. (\ref{atob}) must be generalized so as 
to include  both decay rate and viscous contribution: 
\begin{eqnarray}
&&\nabla_{\mu} {\cal T}^{\mu\nu}_{{\rm a}} = - \Gamma\, g^{\nu\alpha} u_{\alpha}\, ( 
p_{\rm a} + \rho_{\rm a}),
\label{Ac}\\
&& \nabla_{\mu} {\cal T}^{\mu\nu}_{{\rm b}} =  \Gamma\, g^{\nu\alpha} u_{\alpha}\, ( p_{\rm b} + \rho_{\rm b}).
\label{Bc}
\end{eqnarray}
In the case of a spatially flat FRW background, 
denoting with  $\overline{\Gamma}$ the homogeneous component
of the decay rate,  Eqs. (\ref{Ac}) and (\ref{Bc}) 
reduce to 
\begin{eqnarray}
&& 
\dot{\rho}_{\rm a} + 3 H ( \rho_{a} + {\cal P}_{\rm a}) + \overline{\Gamma} ( \rho_{\rm a} + p_{\rm a}) =0
\label{Ac1}\\
&& \dot{\rho}_{\rm b} + 3 H ( \rho_{b} + {\cal P}_{\rm b}) -
 \overline{\Gamma} ( \rho_{\rm a} + p_{\rm a}) =0.
 \label{Bc1}
 \end{eqnarray}
 Notice, once more, the difference between the 
 effective pressures (i.e. ${\cal P}_{\rm a}$ and ${\cal P}_{\rm b}$ 
 defined in Eq. (\ref{effP2})) and the inviscid pressures (i.e. $p_{\rm a}$ and $p_{\rm b}$).
 
 In the following sections we are going to study the evolution of the inhomogeneities 
 in the class of background models introduced in the present section. Therefore, while 
 the background quantities will be only function of the time coordinate, the 
 fluctuations will be functions of both time and  space. 
 
 In the present paper the bulk viscous effects will be parametrized in the context of the Eckart approach \cite{EK1} (which is the one also 
 followed in \cite{weinberg}). It must be mentioned, that this approach is 
 phenomenological in the sense that the bulk viscosity is not 
 modeled on the basis of a suitable microscopic theory.  For  
 caveats concerning the Eckart approach see \cite{isr} (see also 
 \cite{maartens2} and references therein).
 This Eckart approach, however, fits with the phenomenological inclusion 
 of a fluid decay rate that has been also considered 
 recently for related applications to cosmological perturbation theory.
 Possible generalizations of the considerations reported in this paper 
 to causal thermodynamics will not be examined here (see 
 \cite{maa2} for a covariant approach to this problem).

\renewcommand{\theequation}{3.\arabic{equation}}
\section{Cosmological perturbations for imperfect fluids}
\setcounter{equation}{0}
As is well known \cite{bardeen}, density and metric 
perturbations may obey formally different evolution equations,
depending upon the specific gauge used to perform a given calculation. 
 Once a specific gauge-invariant 
quantity is computed in a given gauge, it will be, by definition, the same 
in all the other coordinate systems. Not all coordinate 
systems are equivalent  for specific applications, 
since  one particular gauge may turn out to be more convenient for practical 
reasons.
 
In the following the evolution equations of the fluctuations will 
be derived without choosing a specific gauge. This 
strategy has a twofold advantage. Firstly 
the full gauge-invariant evolution equations can be swiftly derived.
Secondly, the most useful gauge choices 
can be selected by comparing the different forms of the equations 
in different coordinate systems. 
To begin with, the scalar modes  of the Einstein 
equations  will be first derived. We will the move to the evolution 
equations of the hydrodynamical quantities, i.e. the pressure 
and density fluctuations and the peculiar velocity
of each fluid. While in the present and in the following 
sections the attention will be focused on the scalar modes of the geometry, 
some considerations on the evolution of the vector modes are 
collected in the appendix.

\subsection{Einstein system}
The scalar perturbation of Einstein equations can be formally written as 
\begin{equation}
\delta_{\rm s}\, {\cal G}_{\mu}^{\nu} = 8\pi G \,\delta_{\rm s}\, {\cal T}_{\mu}^{\nu},
\label{PE1}
\end{equation}
where $ \delta_{\rm s}$ denotes the first-order scalar fluctuation of a given 
quantity and where ${\cal G}_{\mu}^{\nu}$ denotes the Einstein tensor.
Recalling Eq. (\ref{BV1}) the scalar fluctuation of the effective energy-momentum 
tensor of the fluid sources then becomes,  in general coordinates 
\begin{eqnarray}
\delta_{\rm s} {\cal T}_{\mu}^{\nu} &=& \overline{u}_{\mu}\,\overline{u}^{\nu} [ 
(\delta p + \delta\rho) - \delta \xi\, \overline{\nabla}_{\lambda} \overline{u}^{\lambda} - 
\overline{\xi}\,\,\delta(\nabla_{\lambda}  u^{\lambda})]
\nonumber\\
&+& \delta_{\mu}^{\nu}\,[ \delta \xi \,\overline{\nabla}_{\lambda} \overline{u}^{\lambda} - \delta p + 
\overline{\xi}\,\,\delta(\nabla_{\lambda}  u^{\lambda})]
\nonumber\\
&+& (\delta u_{\mu} \,\overline{u}^{\nu} + \overline{u}_{\mu}\, 
\delta u^{\nu} ) [ ( p + \rho)- 
\overline{\xi} \,\overline{\nabla}_{\lambda} \overline{u}^{\lambda}],
\label{deltacalT}
\end{eqnarray}
where the (total) velocity field satisfies $ g^{\mu\nu}u_{\mu} u_{\nu} =1$
and where the bar means that the corresponding quantity has to be evaluated
using the background geometry. To write Eq. (\ref{deltacalT}) 
in explicit terms one has to make explicit the first-order fluctuation
of the covariant derivative of the velocity field, i.e. 
$\delta (\nabla_{\lambda} u^{\lambda})$. Recalling, in fact, that 
$\nabla_{\lambda} u^{\lambda} = \partial_{\lambda} u^{\lambda} + 
\Gamma_{\beta\alpha} u^{\alpha}$ we have that
\begin{equation}
\delta(\nabla_{\lambda}  u^{\lambda}) = 
\partial_{\lambda} \delta\, u^{\lambda} + \delta \Gamma^{\beta}_{\alpha \beta}
 \overline{u}^{\alpha} + \overline{\Gamma}^{\beta}_{\alpha\beta}
\delta u^{\alpha},
\label{PT}
\end{equation}
 where $\overline{\Gamma}_{\beta\alpha}^{\beta}$ and 
 $\delta \Gamma_{\beta\alpha}^{\beta}$ are, respectively, the 
 background and the perturbed Christoffel connections.
 
Without performing any  gauge choice, 
the scalar fluctuations of  a conformally flat metric with line 
element 
\begin{equation}
ds^2 = \overline{g}_{\mu\nu} dx^{\mu} dx^{\nu} \equiv  a^2(\tau)[ d\tau^2 - d\vec{x}^2],
\label{bm}
\end{equation}
 are  parametrized by four independent 
functions so that the entries of the perturbed metric can be 
written as 
\begin{eqnarray}
&& \delta_{\rm s} g_{00} = 2 a^2 \phi,\,\,\,\,\,\,\,\,\,\,\,\,\,\,\,\,\,\,\,\,\,\,\,\,\,\,\,\,\,\,\,\,\,\,
\,\,\,\,
\delta_{\rm s} g^{00} = - \frac{2}{a^2} \phi
\nonumber\\
&&  \delta_{\rm s} g_{ij} = 
2 a^2 (\psi \delta_{i j} - \partial_{i} \partial_{j} E),\,\,\,\,\,
\delta_{\rm s} g^{ij} = 
-\frac{2}{ a^2} (\psi \delta^{i j} - \partial^{i} \partial^{j} E),
\nonumber\\
&&  \delta_{\rm s} g_{0i} = - a^2  \partial_{i} B,\,\,\,\,\,\,\,\,\,\,\,\,\,\,\,\,\,\,\,\,\,\,\,\,\,\,\,\,
\,\,\,\,\,
\delta_{\rm s} g^{0i} = - \frac{1}{a^2} \partial^{i} B;
\label{SF}
\end{eqnarray}
note that $\delta_{ij}$ is nothing but the Kroneker's delta.

In Eq. (\ref{SF}) the conformal time parametrization of the 
background geometry has been assumed and  the evolution equations of the 
background can be written as  \footnote{We denote 
by a prime a derivation with respect to the conformal time coordinate $\tau$.}
\begin{eqnarray}
&& {\cal H}^2 = \frac{8\pi G}{3} a^2 \rho,
\label{Con1}\\
&& {\cal H}^2 - {\cal H}' = 4\pi G a^2 ( \rho + {\cal P}),
\label{Con2}\\
&& \rho_{\rm a}' + 3 {\cal H} ( \rho_{\rm a} + {\cal P}_{\rm a}) + a \overline{\Gamma} (\rho_{\rm a} + p_{\rm a}) =0,
\label{Con3}\\
&& \rho_{\rm b}' + 3 {\cal H} ( \rho_{\rm b} + {\cal P}_{\rm b}) - a \overline{\Gamma} (\rho_{\rm a} + p_{\rm a}) =0,
\label{Con4}
\end{eqnarray}
where ${\cal H} = a'/a$. In Eqs. (\ref{Con2})--(\ref{Con4}) 
the definition of the total and partial shifted pressures follows from 
Eqs. (\ref{effP}) and (\ref{effP2}).
Since $g^{\mu\nu} u_{\mu}u_{\nu} =1$, it also follows, from  Eqs. (\ref{bm}) 
and (\ref{SF}) that 
\begin{equation}
\overline{u}_{0} = a,\,\,\,\,\,\,\,\,\,\,\,\,\,\,\delta u^{0} = - \phi/a.
\end{equation}

Equations (\ref{PT}) and (\ref{SF}) lead to the following explicit expressions
of the fluctuations of the effective energy-momentum tensor
\begin{eqnarray}
 \delta {\cal T}_{0}^{0} &=& \delta \rho,
 \label{T00}\\
 \delta {\cal T}_{0}^{i} &=&  \biggl( \rho + p - 3 \frac{\overline{\xi}}{a} {\cal H} \biggr) v^{i},
 \label{T0i}\\
 \delta {\cal T}_{i}^{j} &=& - \delta_{i}^{j}\biggl\{ \delta p - 
 3 \frac{{\cal H}}{a} \delta \xi - \frac{\overline{\xi}}{a} [ \theta - 3 
 ( \psi' + {\cal H} \phi) + \nabla^2 E']\biggr\},
 \label{PT2}
 \end{eqnarray}
 where the following notation
 for the peculiar velocity field and for its divergence has been chosen:
 \begin{equation}
 u_{0} \delta u^{i} = v^{i},\,\,\,\,\,\,\,\,\,\,\, \partial_{i} v^{i} = \theta.
 \label{DIV}
 \end{equation}
 Since no specific gauge foxing has been invoked in Eqs. (\ref{T00})--(\ref{PT2}),  
 Eq. (\ref{PE1}) allows the determination of the Hamiltonian constraint
\begin{equation}
 \nabla^2 \psi - 
{\cal H} \nabla^2( B - E') - 3 {\cal H} ( \psi' + {\cal H} \phi) = 4\pi G a^2 \delta \rho,
\label{PE00}
\end{equation}
and the momentum constraint
\begin{equation}
 \nabla^2 ( \psi' + {\cal H} \phi) + ( {\cal H}^2 - {\cal H}') (\nabla^2 B  + \theta) =0.
\label{PE0i}
\end{equation}

In the perturbed geometry (\ref{SF}), the 
 $(i=j)$ and $(i\neq j)$ components of Eq. (\ref{PE1}) become, respectively 
\begin{eqnarray}
&&  \psi''  +({\cal H}^2 + 2 {\cal H}') \phi + {\cal H} (\phi' + 2 \psi') + 
\frac{1}{2}
\nabla^2 [ ( \phi - \psi) + ( B - E')' + 2 {\cal H} ( B - E') ]
\nonumber\\
&&= 4\pi G a^2\biggl\{ \delta p - 3 \frac{ {\cal H}}{a} \delta \xi - \frac{\overline{\xi}}{a}  
[ \theta - 3 (\psi' + {\cal H} \phi) + \nabla^2 E']\biggr\},
\label{Pij}\\
&& - \frac{1}{a^2}
\partial_{i}\partial^{j} [ ( E' - B)' + 2 {\cal H} ( E' - B) + 
(\psi - \phi) ] =0.
\label{Pineqj}
\end{eqnarray}
In Eq. (\ref{Pineqj}) the possible contribution of the 
anisotropic stress has been neglected.

For infinitesimal coordinate transformations, the fluctuations 
of both the metric and the effective energy-momentum 
tensor change. Hence, it is always convenient to define 
appropriate gauge-invariant quantities. A possible 
set of gauge-invariant quantities is then 
\begin{eqnarray}
&& \Phi = \phi + ( B - E')' + {\cal H} ( B - E') 
\label{PHI}\\
&& \Psi= \psi - {\cal H} ( B - E') 
\label{PSI}
\end{eqnarray}
for the metric and 
\begin{eqnarray}
&& \delta \rho_{\rm g} = \delta \rho + \rho' ( B - E')
\label{RHO}\\
&& \delta p_{\rm g} = \delta p + p' ( B - E') 
\label{P}\\
&& \Theta = \theta + \nabla^2 E'
\label{THETA}
\end{eqnarray}
for the (total) fluid sources.  In Eqs. (\ref{RHO}) and (\ref{P}) 
the subscript ``${\rm g}$'' means that the corresponding 
quantity is invariant under infinitesimal coordinate transformations, i.e., 
for short,  gauge-invariant.
This notation will also be employed, when needed, in the following sections.
Equations (\ref{PHI}) and (\ref{PSI}) are the gauge-invariant fluctuations of the metric sometimes called Bardeen potentials 
\cite{bardeen}.  From Eqs. (\ref{PHI}) and (\ref{PSI}), it is clear that 
in the gauge $E=0$ and $B=0$ $\Phi$ and $\Psi$ coincide with 
the longitudinal fluctuations of the metric, i.e. $\phi$ and $\psi$ \cite{bb}.

Since the bulk viscosity 
coefficient is general time dependent, its 
gauge-invariant fluctuation is defined as 
\begin{equation}
\Xi = \delta \xi + \overline{\xi}' ( B - E').
\label{XI}
\end{equation}
Following the usual procedure, the various fluctuations 
both of the geometry and of the sources, i.e. $\psi$, $\phi$, $\delta\rho$ 
, $\delta \xi$ and so on, can be expressed in terms of their 
gauge-invariant counterparts, i.e. $\Psi$, $\Phi$, $\delta\rho_{\rm g}$, 
$\Xi$ and so on. These expressions can then be inserted 
into Eqs. (\ref{PE00}),(\ref{PE0i}) and (\ref{Pij}),(\ref{Pineqj}). 
The resulting set of equations will now contain 
only gauge-invariant quantities. The full system will then be formed by 
the gauge-invariant counterpart of Eqs. (\ref{PE00}) and (\ref{PE0i}):
\begin{eqnarray}
&& \nabla^2 \Psi - 3 {\cal H} ( \Psi' + {\cal H} \Phi) = 
4\pi G a^2 \delta\rho_{\rm g},
\label{G00}\\
&& \nabla^2 ( \Psi' + {\cal H} \Phi) + ({\cal H}^2 - {\cal H}')\Theta = 0 , 
\label{G0i}
\end{eqnarray}
and by the gauge-invariant counterpart of  Eq. (\ref{Pij}):
\begin{equation}
\Psi'' + {\cal H} ( \Phi' + 2 \Psi') + ( {\cal H}^2 + 2 {\cal H}') \Phi = 4\pi G a^2 
\biggl\{ \delta p_{\rm g} - \frac{3 {\cal H}}{a} \Xi - \frac{\overline{\xi}}{a}[ \Theta - 3 ( \Psi' + 
{\cal H} \Phi)]\biggr\}.
\label{Gij}
\end{equation}
 Equation (\ref{Pineqj}) implies 
that $\Phi = \Psi$ in the absence of anisotropic stress.

\subsection{Interacting viscous mixtures}
The perturbed form of the covariant conservation equations can be written as 
\begin{eqnarray}
 \delta(\nabla_{\mu}  {\cal T}^{\mu\nu}_{({\rm a})}) &=& - [ \delta \Gamma \,
\overline{g}^{\nu\alpha} \,\overline{u}_{\alpha}\, ( p_{\rm a} + \rho_{\rm a}) + 
\overline{\Gamma}\, \delta g^{\nu\alpha} \,\overline{u}_{\alpha}\, 
( p_{\rm a} + \rho_{\rm a})
\nonumber\\
&+& \overline{\Gamma} \,\overline{g}^{\nu\alpha} \,\delta u_{\alpha} 
( p_{\rm a} + \rho_{\rm a}) + \overline{\Gamma} \,\overline{g}^{\nu\alpha}\, \overline{u}_{\alpha} ( \delta p_{\rm a } + \delta\rho_{\rm a})],
\label{TA}\\
\delta(\nabla_{\mu} {\cal T}^{\mu\nu}_{({\rm b})}) &=&  [ \delta \Gamma\, 
\overline{g}^{\nu\alpha}\, \overline{u}_{\alpha}\, ( p_{\rm a} + \rho_{\rm a}) + 
\overline{\Gamma}\,\delta g^{\nu\alpha}\, \overline{u}_{\alpha} \,( p_{\rm a}+
\rho_{\rm a}) 
\nonumber\\
&+& \overline{\Gamma}\, \overline{g}^{\nu\alpha} \,\delta u_{\alpha} 
( p_{\rm a} + \rho_{\rm a}) + \overline{\Gamma}\, \overline{g}^{\nu\alpha}\, \overline{u}_{\alpha} \,( \delta p_{\rm a } + \delta\rho_{\rm a})\}.
\label{TB}
\end{eqnarray}
In \cite{KS} the authors choose to parametrize the rate 
of momentum transfer (in the perturbation equations) 
through a more generic functions 
whose properties follow, in the present approach, 
from the form of the coupling already discussed after 
Eqs. (\ref{atob}) (see, for instance, Eq. (5.18) of \cite{KS}).

From Eq. (\ref{TA}), without fixing the gauge, we obtain the following 
general expression from the $(0)$ component
\begin{eqnarray}
&&\delta \rho_{\rm a}' + 3 {\cal H}\,(\delta \rho_{\rm a} + \delta p_{\rm a}) - 
9 \frac{{\cal H}^2}{a}\, \delta \xi_{\rm a} + ( \rho_{\rm a} + {\cal P}_{\rm a})\, 
\theta_{\rm a} - 3\, ( \rho_{\rm a} + {\cal P}_{\rm a})\, \psi' 
\nonumber\\
&&- 3\, \frac{\overline{\xi}_{\rm a}}{a}\, {\cal H}[ \theta_{\rm a} - 3 (\psi' + {\cal H}
\phi) + \nabla^2 E'] + (\rho_{\rm a} + {\cal P}_{\rm a})\, \nabla^2 E' 
\nonumber\\
&& = 
- a [ \delta \Gamma\, ( \rho_{\rm a } + p_{\rm a}) + \phi\, \overline{\Gamma}
(\rho_{\rm a} + p_{\rm a}) + \overline{\Gamma}\,( \delta\rho_{\rm a} 
+\delta p_{\rm a})],
\label{DRa}
\end{eqnarray}
while from Eq. (\ref{TB}), still for the $(0)$ component, the result will be 
\begin{eqnarray}
&&\delta \rho_{\rm b}' + 3 {\cal H}\,(\delta \rho_{\rm b} + \delta p_{\rm b}) - 
9 \frac{{\cal H}^2}{a} \,\delta \xi_{\rm b} \,+ ( \rho_{\rm b} + {\cal P}_{\rm b}) \,
\theta_{\rm b} - 3 \,( \rho_{\rm b} + {\cal P}_{\rm b}) \,\psi' 
\nonumber\\
&&- 3\, \frac{\overline{\xi}_{\rm b}}{a}\, {\cal H}\,[ \theta_{\rm b} - 3\, (\psi' + {\cal H}
\phi) + \nabla^2 E'] + (\rho_{\rm b} + {\cal P}_{\rm b}) \,\nabla^2 E' 
\nonumber\\
&& = 
a [ \delta \Gamma \,( \rho_{\rm a } + p_{\rm a}) + \phi\, \overline{\Gamma}\,
(\rho_{\rm a} + p_{\rm a}) + \overline{\Gamma}\,( \delta\rho_{\rm a} 
+\delta p_{\rm a})].
\label{DRb}
\end{eqnarray}
To derive Eqs. (\ref{DRa}) and (\ref{DRb}) from  Eqs. (\ref{TA}) and (\ref{TB}), 
Eqs. (\ref{Con3}) and (\ref{Con4}) have been repeatedly used.

The $(i)$ component of Eqs. (\ref{TA}) and (\ref{TB}) leads, respectively, to the 
evolution of the peculiar velocity of the a-fluid:
\begin{eqnarray}
&& \theta_{\rm a}' + \biggl[ 4 {\cal H} + \frac{\rho_{\rm a}' + {\cal P}_{\rm a}'}{\rho_{\rm a} 
+ {\cal P}_{\rm a} }
 \biggr] \,\theta_{\rm a} + \nabla^2 (B' +\phi)+ 
 \biggl[ {\cal H} + \frac{{\cal P}_{\rm a}'}{\rho_{\rm a} + {\cal P}_{\rm a}}  
 \biggr] \nabla^2 B
+ \frac{\nabla^2 \,\delta p_{\rm a}}{\rho_{\rm a} + {\cal P}_{\rm a}}
\nonumber\\
&&- \frac{3 {\cal H}  \nabla^2 \delta \xi_{\rm a} }{a\,( \rho_{\rm a} + {\cal P}_{\rm a})} -
\frac{\overline{\xi}_{\rm a} \, [ \nabla^2 \theta_{\rm a} - 3 \nabla^2 ( \psi' + {\cal H}\phi) 
+ \nabla^4 E']}{a (\rho_{\rm a} + {\cal P}_{\rm a})}
 = - a\,\overline{\Gamma} \,
\frac{ \rho_{\rm a} + p_{\rm a} }{\rho_{\rm a} + 
{\cal P}_{\rm a}} \theta,
\label{THa}
\end{eqnarray}
and to the evolution of the peculiar velocity of the b-fluid:
\begin{eqnarray}
&& \theta_{\rm b}' + \biggl[ 4 {\cal H} + \frac{\rho_{\rm b}' + {\cal P}_{\rm b}'}{\rho_{\rm b} 
+ {\cal P}_{\rm b} }
 \biggr] \,\theta_{\rm b} + \nabla^2 (B' +\phi)+ 
 \biggl[ {\cal H} + \frac{{\cal P}_{\rm b}'}{\rho_{\rm b} + {\cal P}_{\rm b}}  
 \biggr] \nabla^2 B
+ \frac{\nabla^2 \,\delta p_{\rm b}}{\rho_{\rm b} + {\cal P}_{\rm b}}
\nonumber\\
&&- \frac{3 {\cal H}  \nabla^2 \delta \xi_{\rm b} }{a\,( \rho_{\rm b} + {\cal P}_{\rm b})} -
\frac{\overline{\xi}_{\rm b} \, [ \nabla^2 \theta_{\rm b} - 3 \nabla^2 ( \psi' + {\cal H}\phi) 
+ \nabla^4 E']}{a (\rho_{\rm b} + {\cal P}_{\rm b})}
 =  a\,\overline{\Gamma} \,
\frac{\rho_{\rm a} + p_{\rm a} }{\rho_{\rm b} + 
{\cal P}_{\rm b}} \theta.
\label{THb}
\end{eqnarray}
Equations (\ref{DRa}), (\ref{DRb}) and (\ref{THa}), (\ref{THb}) can easily be
written only in terms of quantities that are explicitly gauge-invariant. To this 
purpose it is useful to point out that the gauge-invariant fluctuations 
related to the bulk viscosity coefficients and to the decay rate can be 
written as 
\begin{eqnarray}
&&\Xi_{\rm (a,b)} = \delta \xi_{\rm (a,b)} + \overline{\xi}_{\rm (a,b)}' ( B - E'),
\label{XIab}\\
&& \delta \Gamma_{\rm g} = \delta \Gamma + \overline{\Gamma}' ( B - E'),
\label{GAMMA}
\end{eqnarray}
where the subscripts in Eq. (\ref{XIab}) imply that the gauge-invariant 
fluctuation has the same form for both the a- and the b-fluid.

Using Eqs. (\ref{XIab}) and (\ref{GAMMA}),  Eqs. (\ref{DRa}), (\ref{DRb}) 
and (\ref{THa}), (\ref{THb}) can be written in fully gauge-invariant terms. 
More specifically, Eqs. (\ref{DRa}) and (\ref{DRb}) lead, respectively, to 
\begin{eqnarray}
&& \delta \rho_{\rm g\,a}' + 3 {\cal H} ( \delta \rho_{{\rm g\, a}} + \delta p_{{\rm g\,a}} ) - 
9 \frac{{\cal H}^2}{a} \Xi_{\rm a} + ( \rho_{\rm a} + {\cal P}_{\rm a}) \Theta_{\rm a} 
- 3 ( \rho_{\rm a} + {\cal P}_{\rm a}) \Psi' 
\nonumber\\
&&- \frac{3}{a} {\cal H} \overline{\xi}_{\rm a} [ 
\Theta_{\rm a} - 3 ( \Psi' + {\cal H} \Phi)] = 
\nonumber\\
&&- a [ \delta \Gamma_{\rm g}( \rho_{\rm a} + p_{\rm a})  + 
\overline{\Gamma} ( \delta \rho_{\rm g\, a} + \delta p_{\rm g\, a}) + \overline{\Gamma}
( p_{\rm a} + \rho_{\rm a}) \Phi] 
\label{DRGIa}
\end{eqnarray}
and to
\begin{eqnarray}
&& \delta \rho_{\rm g\,b} '+ 3 {\cal H} ( \delta \rho_{{\rm g\, b}} + \delta p_{{\rm g\,b}} ) - 
9 \frac{{\cal H}^2}{a} \Xi_{\rm b} + ( \rho_{\rm b} + {\cal P}_{\rm b}) \Theta_{\rm b} - 
3 ( \rho_{\rm b} + {\cal P}_{\rm b}) \Psi' 
\nonumber\\
&&- \frac{3}{a} {\cal H} \overline{\xi}_{\rm b} [ 
\Theta_{\rm b} - 3 ( \Psi' + {\cal H} \Phi)] = 
\nonumber\\
&&a [ \delta \Gamma_{\rm g}(\rho_{\rm a} + p_{\rm a})   + 
\overline{\Gamma} ( \delta \rho_{\rm g\, a} + \delta p_{\rm g\, a}) + \overline{\Gamma}
( p_{\rm a} + \rho_{\rm a}) \Phi],
\label{DRGIb}
\end{eqnarray}
while Eqs. (\ref{THa}) and (\ref{THb}) become 
\begin{eqnarray}
&&  \Theta_{\rm a}' + \biggl[ 4 {\cal H} +   
\frac{\rho_{\rm a}' + {\cal P}_{\rm a}' }{\rho_{\rm a} + {\cal P}_{\rm a}}
\biggr]\Theta_{\rm a} + \frac{\nabla^2 \delta p_{\rm g\, a}}{\rho_{\rm a} + 
{\cal P}_{\rm a}}
+  \nabla^2 \Phi
\nonumber\\
&&
 - 3 \frac{{\cal H}}{a (\rho_{\rm a} + {\cal P}_{\rm a})} \nabla^2 \Xi_{\rm a} 
- \frac{ \overline{\xi}_{\rm a}[ \nabla^2 \Theta_{\rm a} - 3 \nabla^2(\Psi' + {\cal H} \Phi)]}{a 
(\rho_{\rm a } + {\cal P}_{\rm a})}= 
- a \overline{\Gamma} \frac{\rho_{\rm a} + p_{\rm a}}{\rho_{\rm a} + {\cal P}_{\rm a}} \Theta,
\label{THGIa}\\
&&  \Theta_{\rm b}' + \biggl[ 4 {\cal H} +   
\frac{\rho_{\rm b}' + {\cal P}_{\rm b}' }{\rho_{\rm b} + {\cal P}_{\rm b}}
\biggr]\Theta_{\rm b} + \frac{\nabla^2 \delta p_{\rm g\, b}}{\rho_{\rm b} + 
{\cal P}_{\rm b}}
+  \nabla^2 \Phi
\nonumber\\
&&
 - 3 \frac{{\cal H}}{a (\rho_{\rm b} + {\cal P}_{\rm b})} \nabla^2 \Xi_{\rm b} 
- \frac{ \overline{\xi}_{\rm b}[ \nabla^2 \Theta_{\rm b} - 3 \nabla^2(\Psi' + {\cal H} \Phi)]}{a 
(\rho_{\rm a } + {\cal P}_{\rm b})}= 
 a \overline{\Gamma} \frac{\rho_{\rm a} + p_{\rm a}}{\rho_{\rm b} + {\cal P}_{\rm b}} \Theta.
\label{THGIb}
\end{eqnarray}
In Eqs. (\ref{DRGIa}), (\ref{DRGIb}) and (\ref{THGIa}), (\ref{THGIb}),
$\delta \rho_{\rm  a\,g}$ and $\delta \rho_{\rm b\,g}$ denote 
the gauge-invariant energy density fluctuations of the two species. A similar 
notation has been employed for the fluctuations in the pressure 
densities (i.e. $ \delta p_{\rm a\, g}$ and  $ \delta p_{\rm b\, g}$) 
and for the gauge-invariant fluctuations of the velocity field
(i.e. $\Theta_{\rm a}$ and $\Theta_{\rm b}$). A useful remark is that 
Eqs. (\ref{THGIa}) and (\ref{THGIb}) imply a transfer of momentum 
between the a-fluid and the b-fluid.

The total gauge-invariant fluctuations are easily related to the 
gauge-invariant fluctuations of the single species, i.e. 
\begin{eqnarray}
&&\delta \rho_{\rm g} = \delta \rho_{\rm g\, a} + \delta \rho_{\rm g\, b},
\label{TOTr}\\
&& \delta p_{\rm g} = \delta p_{\rm g\, a} + \delta p_{\rm g\, b},
\label{TOTp}\\
&& (\rho + p) \Theta = (\rho_{\rm a} + p_{\rm a}) \Theta_{\rm a} + 
(\rho_{\rm b} + p_{\rm b}) \Theta_{\rm b},
\label{TOTth}\\
&& \overline{\xi} \Theta = \overline{\xi}_{\rm a} \Theta_{\rm a} + 
\overline{\xi}_{\rm b} \Theta_{\rm b}, 
\label{TOTxi1}\\
&& \Xi = \Xi_{\rm a} + \Xi_{\rm b}.
\label{TOTxi2}
\end{eqnarray}
From Eqs. (\ref{TOTth}) and (\ref{TOTxi1}), it follows that the 
total velocity field can also be written as 
\begin{equation}
(\rho + {\cal P}) \Theta = 
(\rho_{\rm a} + {\cal P}_{\rm a}) \Theta_{\rm a} + (\rho_{\rm b} + {\cal P}_{\rm b}) \Theta_{\rm b},
\label{TOTxi3}
\end{equation}
 where, we recall, ${\cal P} = {\cal P}_{\rm a} + {\cal P}_{\rm b}$.

It is clear that summing up Eq. (\ref{DRGIa}) and Eq. (\ref{DRGIb}), 
the temporal component of the covariant conservation equation 
for the total fluid is recovered, i.e. 
\begin{eqnarray}
&& \delta \rho_{\rm g}'+ 3 {\cal H} ( \delta \rho_{\rm g} + \delta p_{\rm g} ) - 
9 \frac{{\cal H}}{a} \Xi + ( \rho_{\rm a} + {\cal P}_{\rm a}) \Theta
\nonumber\\
&&- 
3 ( \rho+ {\cal P}) \Psi' - \frac{3}{a} {\cal H} \overline{\xi} [ 
\Theta - 3 ( \Psi' + {\cal H} \Phi)] = 0.
\label{DRGITOT}
\end{eqnarray}
Similarly, summing up Eq. (\ref{THGIa}) and Eq. (\ref{THGIb}), the spatial 
component of the covariant conservation equation for the total fluid 
becomes
\begin{eqnarray}
&&  \Theta' + \biggl[  
4 {\cal H} + \frac{( \rho' + {\cal P}')}{ \rho + {\cal P}}\biggr]\Theta+ 
\frac{\nabla^2 \delta p_{\rm g}}{\rho + {\cal P}}+ \nabla^2 \Phi 
\nonumber\\
&&
- 3 \frac{{\cal H}}{a (\rho + {\cal P}) } \nabla^2 \Xi
- \frac{ \overline{\xi} [ \nabla^2 \Theta - 3 \nabla^2(\Psi' + {\cal H} \Phi)]}{\rho + {\cal P}}= 0.
\label{THGITOT}
\end{eqnarray}
As a final comment, it could be noted that while, 
in the case of the background, the effect of the bulk viscosity 
coefficient is to renormalize the effective pressure densities, at the level 
of the fluctuations this simplification is no longer possible as 
is evident, for instance, from Eq. (\ref{Gij}).
In a complementary perspective, the presence of spatial fluctuations 
in the bulk viscosity coefficient induces some sort of intrinsic 
non-adiabatic pressure density variation. 
Each of the fluids of the mixture will have an inviscid contribution and 
a viscous  contribution.  In the limits $\overline{\xi}_{\rm a, b} \to 0$ 
and $\Xi_{\rm a,b}\to 0$, the fluid becomes inviscid both at the level of 
the background and at that of the  fluctuations. 
The inviscid contribution can be characterized, in the simplest case, 
by two barotropic indices and by the two sound speeds, i.e. 
\begin{equation}
w_{\rm a, b} = \frac{p_{\rm a,\, b}}{\rho_{\rm a,\, b}}, \,\,\,\,\,\,\,\,\,
c^2_{\rm s\,a} = \frac{\delta p_{\rm g\, a}}{\delta \rho_{\rm g\, a}}, \,\,\,\,\,\,
c^2_{\rm s\,b} = \frac{\delta p_{\rm g\, b}}{\delta \rho_{\rm g\, b}},
\label{ss0}
\end{equation}
where, in the first equation, the subscript means that the relation holds 
independently for each fluid. 
We are now in a condition to derive the evolution 
equations of the curvature and 
entropy perturbations in the case when both the decay rate and the bulk viscosity 
coefficient experience some spatial variation.

\renewcommand{\theequation}{4.\arabic{equation}}
\section{Curvature and entropy perturbations}
\setcounter{equation}{0}
The curvature 
fluctuations usually discussed in the theory of cosmological 
perturbations \cite{bardeen,bst,bp,lyth} are indeed perturbations of the spatial 
curvature on comoving orthogonal hypersurfaces (see, for instance, \cite{maxrev}).
Gauge-invariant curvature fluctuations can be 
defined, in terms of the Bardeen potentials, as 
\begin{equation}
{\cal R} = - \biggl[\Psi + \frac{{\cal H}}{{\cal H}^2 - {\cal H}'}({\cal H} \Phi + \Psi') \biggr] = - ( \Psi - {\cal H} V_{\rm g}),
\label{defR}
\end{equation}
where, in the second equality, we used the notation
\footnote{In the following paragraphs of the present section 
the notation $\theta = \nabla^2 v$ and $\Theta = \nabla^2 V_{\rm g}$ will 
be employed; these definitions follow naturally from the conventions 
of Eqs. (\ref{DIV}) and (\ref{THETA}).}
 $\Theta = \nabla^2 V_{\rm g}$.
Note that the second equality follows by using the momentum 
constraint.
Suppose, indeed, to be in the comoving orthogonal gauge where, by definition
$(v_{\rm C} + B_{\rm C} )=0$. In this gauge, the fluctuations of the spatial 
curvature are related to $\psi_{\rm C}$. Suppose now to shift by an infinitesimal quantity $\epsilon_{0}$ the conformal time coordinate 
$\tau_{\rm C}$, i.e. 
\begin{equation}
\tau_{\rm C} \to \tau = \tau_{\rm C} + \epsilon_{0}.
\label{shift1}
\end{equation}
Under the coordinate shift (\ref{shift1}),
 both $\psi_{\rm C}$ and $(v_{\rm C} + B_{\rm C})$ transform as 
\begin{eqnarray}
&&(v_{\rm C} + B_{\rm C}) \to v + B = v_{\rm C} + B_{\rm C} + \epsilon_{0},
\label{vBC}\\
&& \psi_{\rm C} \to \psi = \psi_{\rm C} + {\cal H} \epsilon_{0}.
\label{psiC1}
\end{eqnarray}
Consequently, the expression of $\psi_{\rm C}$ in the new coordinate 
system defined by Eq. (\ref{shift1}) is simply 
\begin{equation}
\psi_{\rm  C} = \psi - {\cal H} ( v + B) = (\Psi - {\cal H} V_{\rm g}).
\label{psiC2}
\end{equation}
In the first equality of Eq. (\ref{psiC2}) the value of $\epsilon_{0}$ is fixed by 
Eq. (\ref{vBC}). The second equality of Eq. (\ref{psiC2}) follows, instead, 
from the definition of gauge-invariant fluctuations of Eqs. (\ref{PSI}) 
and (\ref{THETA}). Clearly, up to a sign (this is a matter of conventions) 
Eq. (\ref{psiC2}) is exactly Eq. (\ref{defR}). 
While in  the comoving orthogonal gauge, ${\cal R}$ is related to fluctuations 
of the spatial curvature, in a different coordinate system ${\cal R}$ will have 
the same numerical value (by gauge-invariance) but will not necessarily 
be related to curvature fluctuations.

There is also another  physical quantity that plays an important 
r\^ole in our considerations, namely the curvature fluctuation on uniform 
density hypersurfaces (see, for instance, \cite{maxrev}).  The fluctuations
of the (total) curvature 
on uniform density hypersurfaces have a gauge-invariant interpretation
and they can be defined as 
\begin{equation}
\zeta = - \biggl( \Psi + {\cal H} \frac{\delta \rho_{\rm g}}{\rho'}\biggr).
\label{defzeta}
\end{equation}
As in the case discussed in Eq. (\ref{defR}), $\zeta$ coincides 
with a perturbation of the spatial curvature in the uniform density 
gauge (see, for instance, \cite{maxrev}). In a diffferent 
coordinate system $\zeta$ will keep the same numerical value,
but its physical interpretation will be different. Suppose to be in the 
coordinate system defined by the condition $\delta \rho_{\rm D}$. This 
is the so-called uniform density gauge. Consider then curvature 
fluctuations on the uniform density hypersurfaces $\psi_{\rm D}$. Under 
an infinitesimal shift of the time coordinate, $\psi_{\rm D}$ and 
$\delta \rho_{\rm D}$ change as 
\begin{eqnarray}
&& \psi_{\rm D} \to \psi = \psi_{\rm D} + {\cal H} \epsilon_{0},
\label{psiD1}\\
&& \delta \rho_{\rm D} \to \delta \rho = \delta \rho_{\rm D} - \rho' \epsilon_{0}.
 \label{dRD}
 \end{eqnarray} 
Equations (\ref{psiD1}) and (\ref{dRD}) imply that, in the new coordinate system,
\begin{equation}
\psi_{\rm D} = \psi + {\cal H} \frac{\delta\rho}{\rho'} = \Psi 
+ {\cal H} \frac{\delta\rho_{\rm g}}{\rho'},
\label{psiD2}
\end{equation}
where the first equality follows from Eq. (\ref{dRD}) and the second 
 from Eqs. (\ref{PSI}) and (\ref{RHO}).  Again, up to a
sign, Eq. (\ref{psiD2}) coincides with Eq. (\ref{defzeta}). 
This means that $\zeta$ parametrizes fluctuations in the 
spatial curvature computed on uniform density hypersurfaces. In a 
different gauge the numerical value of $\zeta$ will remain the same but 
its interpretation may be completely different.

By taking the difference of Eqs. (\ref{defR}) and (\ref{defzeta}) and 
by using the Hamiltonian constraint in its gauge-invariant form, i.e. 
Eq. (\ref{G00}), the following relation between ${\cal R}$ and $\zeta$ 
can be obtained:
\begin{equation}
{\cal R} = \zeta - \frac{\nabla^2 \Psi}{12 \pi G a ^2 ( \rho + {\cal P})},
\label{Rtozeta}
\end{equation}
where, as usual, ${\cal P}$ is the total shifted pressure of the mixture.
Equation (\ref{Rtozeta}) implies that, up to Laplacians of $\Psi$, 
${\cal R} \simeq \zeta$. Equation (\ref{Rtozeta}) generalizes the relation 
between ${\cal R}$ and $\zeta$ to the case when the bulk viscosity 
coefficient is non-vanishing.  In fact, in the denominator at the 
right-hand side of Eq. (\ref{Rtozeta}) there appears ${\cal P}$, i.e. the 
sum of inviscid and viscous contributions to the background 
pressure.

As elaborated,  Eq. (\ref{defzeta}) describes 
curvature fluctuations on uniform density hypersurfaces.
In the same spirit, it is also possible to define the curvature fluctuations 
on the hypersurfaces where the energy densities of either the a-fluid 
or the b-fluid are uniform, namely:
\begin{eqnarray}
&&
\zeta_{\rm a} = -\biggl( \Psi + {\cal H}\frac{\delta\rho_{\rm g\,a}}{\rho_{\rm a}'}\biggr),
\label{defzetaa}\\
&& \zeta_{\rm b} =
 -\biggl( \Psi + {\cal H}\frac{\delta\rho_{\rm g\,b}}{\rho_{\rm b}'}\biggr),
\label{defzetab}
\end{eqnarray}
where, according to Eqs. (\ref{DRGIa}) and (\ref{DRGIb}), 
$\delta \rho_{\rm g\,a}$ and $\delta\rho_{\rm g\,b}$ are the gauge-invariant density 
fluctuations of each fluid of the mixture.
In equivalent terms, Eqs. (\ref{defzetaa}) and (\ref{defzetab}) 
can also be interpreted as related to 
 the density fluctuations on the hypersurfaces where
the curvature is homogeneous (this point of view will be particularly useful
for the considerations of the following section). 

From the gauge-invariant definitions given in Eqs. (\ref{defzeta}) and (\ref{defzetaa}), (\ref{defzetab}) it is easy to show that 
\begin{equation}
\zeta = \frac{\rho_{\rm a}'}{\rho'} \zeta_{\rm a} +  \frac{\rho_{\rm b}'}{\rho'} \zeta_{\rm b}.
\label{zetatot}
\end{equation}
While the weighted sum of $\zeta_{\rm a}$ and $\zeta_{\rm b}$ is related to 
$\zeta$, the difference between them is related to 
the entropy fluctuations. More specifically, the relative 
fluctuations in the specific entropy $\varsigma$ can be written as 
\begin{equation}
{\cal S} = \frac{\delta \varsigma}{\varsigma} = -3 (\zeta_{\rm a} - \zeta_{\rm b}).
\label{entropydef}
\end{equation}
In the limit of vanishing decay rate and vanishing bulk viscous 
stresses (for two perfect, inviscid, non-interacting fluids), 
Eq. (\ref{entropydef}) becomes 
\begin{equation}
{\cal S} = \frac{\delta_{\rm g\, a}}{( 1 + w_{\rm a})} -  
\frac{\delta_{\rm g\, b}}{( 1 + w_{\rm b})} ,
\label{KSdef}
\end{equation}
where $\delta_{\rm g\,a} = \delta \rho_{\rm g\, a}/\rho_{\rm a}$ 
and $\delta_{\rm g\, b} = \delta \rho_{\rm g\, b}/\rho_{\rm b}$  
are the gauge-invariant density contrasts; $w_{\rm a}$ and $w_{\rm b}$ 
are the barotropic indices for the two fluids of the mixture. Equation (\ref{KSdef}) 
coincides with the definition of Ref. \cite{KS} and it applies, for instance, in
 the discussion of CDM-radiation isocurvature 
mode in the absence of decay rates.  In  more general situations, for instance 
when decay rates are included, 
Eq. (\ref{entropydef}) does not reduce to Eq. (\ref{KSdef}) \cite{malikwands}. 
Recalling 
Eqs. (\ref{defzetaa}) and (\ref{defzetab}), Eq. (\ref{entropydef})  can 
 also be written as 
\begin{equation}
{\cal S} =    3 {\cal H} \biggl( \frac{\delta\rho_{\rm g\,a}}{\rho_{\rm a}'} - 
\frac{\delta \rho_{\rm g\,b}}{\rho_{\rm b}'}\biggr).
\label{entropydef2}
\end{equation}

Let us now define the quantities relevant to the evolution 
of curvature perturbations. 
The total pressure density appearing in Eq. (\ref{Gij}) 
can always be split as  
\begin{equation}
\delta p_{\rm g} = \biggl(\frac{\delta p_{\rm g}}{\delta \rho_{\rm g}}\biggr)_{\varsigma} \delta\rho_{\rm g}
+ \biggl(\frac{\delta p_{\rm g}}{\delta \varsigma}\biggr)_{\rho} \delta\varsigma,
\label{deltapg}
\end{equation}
where the two subscripts imply that the two relative variations 
at the right-hand side should be taken, respectively, at {\em constant}
entropy and energy densities. Thus, to perform the variation 
at constant (total) energy density means that
$\delta \rho_{\rm g\, a} = - \delta\rho_{\rm g \,b }$. 
Similarly, to perform the variation at constant $\varsigma$ means that 
$\delta\varsigma =0$, i.e. from Eq. (\ref{entropydef2})
\begin{equation}
\frac{\delta\rho_{\rm g\,a}}{\rho_{\rm a}'} =
\frac{\delta \rho_{\rm g\,b}}{\rho_{\rm b}'}.
\end{equation}
Notice also that, for practical purposes, it is often useful to denote 
the whole second term at the right-hand side of Eq. (\ref{deltapg}) as 
$\delta p_{\rm nad}$.

In the case of two fluids the total speed of sound and the non-adiabatic
pressure density variation are:
\begin{eqnarray}
&& c_{\rm s}^2 =
 \biggl( \frac{\delta p_{\rm g}}{\delta \rho_{\rm g}}\biggr)_{\varsigma} = 
 \frac{c^2_{\rm s\, a} \rho_{\rm a}' + c^2_{\rm s\,b} \rho_{\rm b}'}{\rho_{\rm a}' + \rho_{\rm b}'},
 \label{cs2tot}\\
&& \delta p_{\rm nad} =  \biggl( \frac{\delta p}{\delta \varsigma}\biggr)_{\rho} \delta \varsigma = 
- \frac{(c^2_{\rm s\,a} - c^2_{\rm s\,b}) \rho_{\rm a}' \rho_{\rm b}'}{{\cal H} ( \rho_{\rm a}' + 
\rho_{\rm b}')} (\zeta_{\rm a} - \zeta_{\rm b}),
\label{deltapnad1}
\end{eqnarray}
where the speeds of sound in the two fluids of the mixture have been explicitly 
introduced.
Recalling the connection between $\zeta$ and the weighted sum of
$\zeta_{\rm a}$ and $\zeta_{\rm b}$, i.e. Eq. (\ref{zetatot}), it is also possible to write 
$\delta p_{\rm nad}$ in an alternative useful form:
\begin{equation}
\delta p_{\rm nad} =  \frac{(c^2_{\rm s\,b} - c^2_{\rm s\,a}) \rho_{\rm a}' }{{\cal H}} (\zeta_{\rm a} - \zeta),
\label{deltapnad2}
\end{equation}
where, according to Eq. (\ref{ss0}) the speeds of sound refer to the 
inviscid contribution to the total energy-momentum tensor. Thus 
$c^2_{\rm s,\, b} = w_{\rm b}$ and $c^2_{\rm s, a} = w_{\rm a}$.

Taking now the difference between Eqs. (\ref{Gij}) and (\ref{G00}), and recalling 
 Eqs. (\ref{defR}) and  (\ref{deltapg}), 
 the evolution of ${\cal R}$ can be obtained easily, and it turns out to be 
 \begin{eqnarray}
&&{\cal R}' = \frac{ 3 {\cal H}}{a ( \rho + {\cal P})} \overline{\xi}'  ( {\cal R} + \Psi) 
- \frac{{\cal H}}{\rho + {\cal P}} \delta p_{\rm nad} 
+ \frac{3 {\cal H}^2 }{a ( \rho + {\cal P})} \Xi 
\nonumber\\
&&+ \frac{\overline{\xi}}{a}
\frac{{\cal H}}{\rho + {\cal P}} \Theta - \frac{{\cal H} c_{\rm s}^2}{4\pi G a^2(\rho + {\cal P})}
\nabla^2 \Psi.
\label{evolR}
\end{eqnarray}
We can also write the above expression directly in terms 
of $\zeta$, i.e. as the density contrast on uniform 
density hypersurfaces. Recalling Eq. (\ref{Rtozeta}) we can also write 
\begin{eqnarray}
&& {\cal R}' = \zeta'  - \frac{\nabla^2 \Psi'}{12 \pi G a^2 (\rho +{\cal P})} - 
\frac{{\cal H} ( 3 c_{\rm s}^2 + 1)}{12 \pi G a^2 ( \rho + {\cal P})} \nabla^2 \Psi 
\nonumber\\
&& 
- \frac{ {\cal H}}{ 4 \pi G a^2 ( \rho + {\cal P})^2} \frac{\overline{\xi}'}{a} 
\nabla^2 \Psi + \frac{\overline{\xi}}{a (\rho + {\cal P})} \nabla^2 \Psi. 
\label{Rpr}
\end{eqnarray}
by simply taking the conformal-time derivatives of the left- and of the 
right-hand sides.
Inserting Eq. (\ref{Rpr}) into Eq. (\ref{evolR}) and 
using the gauge-invariant form of the momentum constraint 
given in Eq. (\ref{G0i}), we simply obtain
\begin{eqnarray}
&&\zeta' = \frac{ 3 {\cal H}}{a ( \rho + {\cal P})} \overline{\xi}'   \zeta 
- \frac{{\cal H}}{\rho + {\cal P}} \delta p_{\rm nad} 
+ \frac{3 {\cal H} }{a ( \rho + {\cal P})} ( {\cal H} \Xi  + \overline{\xi}' \Psi) 
\nonumber\\
&&+ \Theta\biggl[ \frac{\overline{\xi} {\cal H}}{a ( 
\rho + {\cal P})} - \frac{1}{3}\biggr] -\frac{{\cal H}}{12 \pi G a^2 ( \rho + {\cal P})}\nabla^2 
( \Phi - \Psi) - \frac{\overline{\xi}}{a (\rho + {\cal P})} \nabla^2 \Psi.
\label{evolzeta}
\end{eqnarray}
Concerning Eq. (\ref{evolzeta}), a few comments are in order:
\begin{itemize}
\item{} the first three terms on the right-hand side  of Eq. (\ref{evolzeta}) 
dominate over length scales larger than the Hubble radius;
the  remaining terms  in the second line of Eq. (\ref{evolzeta})
are subleading in this limit;
\item{} when $\overline{\xi}\to 0$ and $\Xi\to 0$ (i.e. in the case of vanishing 
homogeneous and inhomogeneous viscosity), Eq. (\ref{evolzeta}) 
reduces to the usual form of the evolution equation 
for $\zeta$ (note, in fact, that in this limit $ {\cal P} \to p$);
\item{} since $\delta p_{\rm nad}$ contains both $\zeta$ and $\zeta_{\rm a}$ 
(being proportional to the fluctuations of the entropy density), the evolution 
of $\zeta$ is completely specified only when coupled to the evolution of $\zeta_{\rm a}$.
\end{itemize}

Therefore, to have a self-consistent system of differential equations 
for length scales larger than the Hubble radius,
it is mandatory to derive an evolution equation for $\zeta_{\rm a}$. 
This can be deduced by
inserting $\delta \rho_{\rm g\, a}$ (obtained from Eq. (\ref{defzetaa})) into 
Eq. (\ref{DRGIa}). The result of this algebraic operation can then be written as 
\begin{eqnarray}
&& (\zeta_{\rm a} +\Psi)' +  (\zeta_{\rm a} +\Psi) \biggl[ \frac{{\cal H}}{\rho_{\rm a}'} \biggl(\frac{\rho_{\rm a}'}{{\cal H}} \biggr)' + 3 {\cal H}( c_{\rm s\,a } +1) + a \overline{\Gamma} (c_{\rm s\,a} + 1) \biggr] 
\nonumber\\
&& + \frac{9 {\cal H}^3}{a \rho_{\rm a}'}
\Xi_{\rm a} - \frac{{\cal H} ( \rho_{\rm a} + {\cal P}_{\rm a})}{\rho_{\rm a}'} \Theta_{\rm a} 
\nonumber\\
&& \frac{3{\cal  H} (\rho_{\rm a} + {\cal P}_{\rm a})}{\rho_{\rm a}'} \Psi' +
\frac{3}{a} \frac{{\cal H}^2}{\rho_{\rm a}'} \overline{\xi}_{\rm a} \biggl[ 
\Theta_{\rm a} - 3 \frac{({\cal H}' -{\cal H}^2)}{{\cal H}} ({\cal R} +\Psi) \biggr]
\nonumber\\
&& 
= a (p_{\rm a } + \rho_{\rm a}) \overline{\Gamma} \frac{{\cal H}}{\rho_{\rm a}'} 
\biggl[ \delta_{\Gamma_{\rm g}} + \frac{{\cal H}' - {\cal H}^2}{{\cal H}^2} {\cal R}\biggr]
- a \frac{(p_{\rm a} + \rho_{\rm a})}{\rho_{\rm a}'} \overline{\Gamma} \biggl[
\Psi' + \frac{{\cal H}^2 - {\cal H}'}{{\cal H}} \Psi\biggr].
\label{evolzetaa}
\end{eqnarray}
where 
\begin{equation} 
\delta_{\Gamma_{\rm g}} = \frac{\delta \Gamma_{\rm g}}{\overline{\Gamma}}.
\end{equation}
is the fractional fluctuation of the decay rate.  

Equation (\ref{evolzetaa}) can be simplified by using the evolution equations 
of the background and, in particular, the background conservation equation 
for the a-fluid, i.e. Eq. (\ref{Con3}). This observation allows us to eliminate the 
various terms proportional to  $\Psi'$ and to $\Psi$. The second achievable 
simplification is to trade ${\cal R}$ for $\zeta$. This operation, according 
to Eq. (\ref{Rtozeta}), will bring 
some extra terms proportional to $\nabla^2 \Psi$. In the case where 
the inviscid part of the a-fluid has a barotropic index $w_{\rm a}$, the 
 speed of sound is given by $c^2_{\rm s\, a}= w_{\rm a}$. In this case 
the quantity  
\begin{equation}
g_{\rm a} = - \frac{{\cal H} ( 1 + w_{\rm a}) \rho_{\rm a}}{\rho_{\rm a}'},
\label{defga}
\end{equation}
can be defined.

Taking into account all these simplifications, Eq. (\ref{evolzetaa}) can be 
written as 
\begin{eqnarray}
&& \zeta_{\rm a}' + \biggl[ \frac{9 {\cal H}^2}{a} \frac{\overline{\xi}_{\rm a}}{ \rho_{\rm a}} - \frac{g_{\rm a}'}{g_{\rm a}} \biggr] \zeta_{\rm a} + 
\frac{9 {\cal H}^2}{a \rho_{\rm a}'} [ {\cal H} \Xi_{\rm a} + \overline{\xi}_{\rm a}' \Psi ] 
\nonumber\\
&& + \frac{ 3 {\cal H}^2}{a \rho_{\rm a}'} \overline{\xi}_{\rm a} \biggl[ \Theta_{\rm a} 
+ 3 \frac{{\cal H}^2 - {\cal H}'}{{\cal H}} \zeta\biggr] + 3 \frac{{\cal H} \overline{\xi}_{\rm a} }{ a\rho_{\rm a}'} 
\nabla^2\Psi 
\nonumber\\
&& = - a \overline{\Gamma} g_{\rm a} \biggl[\biggl( \delta_{\Gamma}
+ \frac{\overline{\Gamma}'}{\overline{\Gamma} {\cal H}} \Psi\biggr) + \frac{{\cal H}' - {\cal H}^2}{{\cal H}^2} \zeta \biggr] - a \overline{\Gamma}\frac{g_{\rm a}}{3 {\cal H}^2} \nabla^2 \Psi
\label{evolzetaa2}
\end{eqnarray}
ln Eq. (\ref{evolzetaa2}), several terms are 
subleading for length scales larger than the Hubble radius.  
Thus, we can rewrite Eqs. (\ref{evolzeta}) and (\ref{evolzetaa2})
in the cosmic time coordinate, keeping only the dominant 
terms for length scales larger than the Hubble radius. The result
is given by 
\begin{eqnarray}
\dot{\zeta} = \frac{ 3 H}{ ( \rho + {\cal P})} \dot{\overline{\xi}}   \zeta
- \frac{H}{\rho + {\cal P}} \delta p_{\rm nad} 
+ \frac{3 H}{ ( \rho + {\cal P})} ( H \Xi +\dot{\overline{\xi}} \Psi)
\label{zetagi}
\end{eqnarray}
and by 
\begin{eqnarray}
&& \dot{\zeta}_{\rm a} + \biggl[ 9 H^2 \frac{\overline{\xi}_{\rm a}}{\rho_{\rm a}} - \frac{\dot{g}_{\rm a}}{g_{\rm a}} \biggr] \zeta_{\rm a} + 
\frac{9  H^2}{a \dot{\rho}_{\rm a}} [  H\Xi_{\rm a} + \dot{\overline{\xi}}_{\rm a} \Psi ]  - \frac{9  H \dot{H}}{\dot{\rho}_{\rm a}} \overline{\xi}_{\rm a}\zeta  
\nonumber\\
&& = - \overline{\Gamma} g_{\rm a} \biggl[\biggl( \delta_{\Gamma}
+ \frac{\dot{\overline{\Gamma}}}{\overline{\Gamma} 
 H} \Psi\biggr) + \frac{\dot{H}}{ H^2} \zeta \biggr]. 
\label{zetaagi}
\end{eqnarray}
Equations (\ref{zetagi}) and (\ref{zetaagi}) 
can be used, for instance, to study numerically the evolution 
of fluctuations whose wavelength is larger than the Hubble radius, 
with two caveats. The first caveat is that $g_{\rm a}$ has to be always 
non-singular. This may not always be the case 
when a viscous contribution is present.  In this case, instead of studying 
the evolution of $\zeta_{\rm a}$ it will be more practical 
to deal directly with $\delta_{\rm g\, a}$, i.e. the 
gauge-invariant density contrast. The second remark 
is that, apparently, Eqs. (\ref{zetagi}) and (\ref{zetaagi}) 
require the independent determination of $\Psi$. The evolution equation 
for $\Psi$ can be directly obtained in terms of $\zeta$ from Eqs. (\ref{defR}) 
and (\ref{Rtozeta}). 
As  will be shown in the next section, the 
coupled evolution of $\zeta_{\rm a}$ and $\zeta$ can also be easily 
integrated directly in a specific gauge, i.e. the so-called 
off-diagonal gauge \cite{offdiagonal} (see also \cite{hwang2} and \cite{maxrev}).

It is appropriate to remark here that in the case of homogeneous 
decay rate (i.e. $\delta_{\Gamma_{\rm g}} =0$) and in the 
absence of homogeneous and inhomogeneous viscosity 
(i.e. $\overline{\xi}_{\rm a} = \overline{\xi}_{\rm b} =0$ and $\Xi_{\rm a} 
= \Xi_{\rm b} =0$),  Eqs. (\ref{zetagi}) and (\ref{zetaagi}) 
become simply, in units $8\pi G= 1$,
\begin{eqnarray}
&& \dot{\zeta} = \frac{w_{\rm b} - w_{\rm a}}{\rho + p}\, 
\dot{\rho}_{\rm a} \, ( \zeta - \zeta_{\rm a} ),
\label{zsim1}\\
&& \dot{\zeta}_{\rm a}  = \overline{\Gamma} g_{\rm a} 
\frac{\dot{H}}{H^2} (\zeta_{\rm a} - \zeta).
\label{zasim2}
\end{eqnarray}
Equation (\ref{zsim1}) can be deduced from Eq. (\ref{zetagi}) 
using Eq. (\ref{deltapnad1}) in the specific case 
$c_{\rm s,\, a}^2 = w_{\rm a}$ and $c_{\rm s\, b}^2 = w_{\rm b}$.
To get to Eq. (\ref{zsim1}) it is also useful to recall that 
$\zeta_{\rm b}$ can be always eliminated through Eq. (\ref{zetatot}).
Note that in Eq. (\ref{zsim1}), at the right-hand side, 
the denominator reads $(\rho + p)$ and not $(\rho +{\cal P})$ 
since ${\cal P}\to p$ in the limit $\overline{\xi} =0$.
Equation (\ref{zasim2}) can be obtained from Eq. (\ref{zetaagi}) 
by making explicit the expressions depending on $g_{\rm a}$ 
and by using Eq. (\ref{Ac1}) always in the case $c_{\rm s,\, a}^2 = w_{\rm a}$.

\renewcommand{\theequation}{5.\arabic{equation}}
\section{Off-diagonal gauge description}
\setcounter{equation}{0}
In the off-diagonal gauge, the diagonal entries of the perturbed 
metric are set to zero, i.e.  $\delta_{\rm s} g_{ij}=0$ in Eq. (\ref{SF}).
More precisely, in the off-diagonal gauge, 
\begin{equation}
\psi_{\rm od} =0,\,\,\,\,\,\,\,\,\,\,\, E_{\rm od} =0.
\end{equation}
From a generic perturbed line element where all the entries of the 
perturbed metric do not vanish, the 
off-diagonal coordinate system can be reached by performing a coordinate 
transformation 
\begin{eqnarray}
&& \tau \to \tilde{\tau} = \tau + \epsilon_{0},
\nonumber\\
&& x_{i} \to \tilde{x}_{i} = x_{i} + \partial_{i} \lambda.
\label{shiftod}
\end{eqnarray}
The shift (\ref{shiftod}) induces a transformation in the  $\delta_{\rm s} g_{ij}$ 
components of the perturbed metric:
\begin{eqnarray}
&& \psi \to \psi_{\rm od} = \psi + {\cal H} \epsilon_{0},
\nonumber\\
&& E\to E_{\rm od} = E - \lambda.
\label{shift2}
\end{eqnarray}
The gauge parameters are completely fixed to $\lambda = E$ and 
$\epsilon_{0} = - \psi/{\cal H}$ since, as introduced above, 
$\psi_{\rm od}= E_{\rm od}=0$.  In the off-diagonal gauge
the curvature is then uniform, i.e. $\psi_{\rm od}=0$. 
This is the reason why the off-diagonal gauge is also 
correctly named uniform-curvature gauge. 

In the off-diagonal coordinate system, the gauge-invariant variables 
$\zeta_{\rm a}$ and $\zeta_{\rm b}$ defined in the previous section are 
simply proportional to the relative density contrasts of the two fluids, i.e. 
\begin{eqnarray}
&& \zeta_{\rm a} = - {\cal H} \frac{\delta \rho_{\rm a}}{\rho_{\rm a}'} = - 
{\cal H} \frac{\rho_{\rm a}}{\rho_{\rm a}'} \delta_{\rm a} 
\nonumber\\
&&  \zeta_{\rm b} = - {\cal H} \frac{\delta \rho_{\rm b}}{\rho_{\rm b}'} = - 
{\cal H} \frac{\rho_{\rm b}}{\rho_{\rm b}'} \delta_{\rm b},
\label{zetaab}
\end{eqnarray}
where $\delta_{\rm a}$ and $\delta_{\rm b}$ are the density contrasts 
computed in the off-diagonal gauge.

The evolution equations in the off-diagonal gauge can be obtained directly 
from the Eqs. (\ref{PE00}),(\ref{PE0i}) and (\ref{Pij})-,(\ref{Pineqj}) by setting $E$ and $\psi$ to 0. With this strategy in mind
the Hamiltonian and momentum constraints read, respectively \footnote{In the following, without ambiguity, $\phi$, $B$, $\theta$ and all the other perturbation 
variables are meant to be evaluated in the off-diagonal gauge.},
\begin{eqnarray}
&& - {\cal H} \nabla^2 B - 3 {\cal H}^2 \phi = 4\pi G a^2 \delta \rho,
\label{OD1}\\
&& {\cal H} \nabla^2 \phi + ({\cal H}^2 - {\cal H}') \nabla^2 B = - 4 \pi G a^2 (p +\rho) 
\theta,
\label{OD2}
\end{eqnarray}
while the $(i=j)$ and $(i\neq j)$ components
 of the perturbed equations of motion are, respectively,
 \begin{eqnarray}
 && ({\cal H}^2 + 2 {\cal H}') \phi + {\cal H}\phi' = 4\pi G a^2 \delta p,
 \label{OD3}\\
 && B' + 2 {\cal H} B + \phi=0.
 \label{OD4}
 \end{eqnarray}
 
 The covariant conservation of the perturbed energy-momentum tensor 
 implies, instead, 
 \begin{eqnarray}
 && \delta \rho_{\rm a}' + 3 {\cal H}( \delta \rho_{\rm a} + \delta p_{\rm a}) + 
 (\rho_{\rm a} + p_{\rm a}) \theta_{\rm a} =
 - a ( \rho_{\rm a} 
 + p_{\rm a})   \overline{\Gamma} \biggl( \delta_{\Gamma} + \phi  + 
 \frac{\delta\rho_{\rm a} + \delta p_{\rm a}}{\rho_{\rm a} + p_{\rm a}}\biggr),
 \label{OD5}\\
 && \delta \rho_{\rm b}' + 3 {\cal H}( \delta \rho_{\rm b} + \delta p_{\rm b}) + 
 (\rho_{\rm b} + p_{\rm b}) \theta_{\rm b} = 
 a  (p_{\rm a} + \rho_{\rm a}) \overline{\Gamma} \biggl( \delta_{\Gamma} + \phi + 
 \frac{\delta\rho_{\rm a} + \delta p_{\rm a}}{\rho_{\rm a} + p_{\rm a} }\biggr).
\label{OD6}
\end{eqnarray}

In the off-diagonal gauge the explicit form of the evolution equations 
for $\zeta$ and $\zeta_{\rm a}$ (amputated to the leading terms 
that are relevant for length scales larger than the Hubble radius) are:
\begin{equation}
\dot{\zeta} = - \frac{3 H}{2 \dot{H}} [ H (\delta \xi_{\rm a} + \delta\xi_{\rm b}) + (\dot{\overline{\xi}}_{\rm a} + \dot{\overline{\xi}}_{\rm b}) \zeta] 
+ \frac{1}{2 \dot{H}} (w_{\rm b} - w_{\rm a}) \dot{\rho}_{\rm a} ( \zeta_{\rm a} 
-\zeta),
\label{zetaex1}
\end{equation}
where we traded $(\rho + {\cal P})$ for $- 2 \dot{H}$, as  follows from Eq. (\ref{B2}) 
(when units $8\pi G = 1$ are used \footnote{In the following, without loss of generality, we will set $8\pi G = 1$.}).  The evolution equation 
for $\zeta_{\rm a}$ is then 
\begin{eqnarray}
\dot{\zeta}_{\rm a} + \biggl( 9 H^2 \frac{\overline{\xi}_{\rm a}}{\rho_{\rm a}} - \frac{\dot{g}_{\rm a}}{g_{\rm a}} \biggr) \zeta_{\rm a} + 
\frac{9  H}{\dot{\rho}_{\rm a}} (  H^2 \delta\xi_{\rm a} -    \dot{H}\overline{\xi}_{\rm a}\zeta  )= - 
\overline{\Gamma} g_{\rm a} \biggl( \delta_{\Gamma}
 + \frac{\dot{H}}{ H^2} \zeta \biggr) .
\label{zetaaex1}
\end{eqnarray}
To derive Eqs. (\ref{zetaex1}) and (\ref{zetaaex1}) directly in the 
off-diagonal gauge, it is useful to notice that, in this coordinate 
system, the relation of $\zeta$ to $\phi$ is particularly simple and it is given by 
\begin{equation}
\zeta = - \frac{{\cal H}^2}{{\cal H}^2 - {\cal H}'} \phi \equiv  \frac{H^2}{\dot{H}} \phi.
\end{equation}
 As previously discussed, it is useful for some applications to study curvature fluctuations directly in the $\zeta$, $\delta_{\rm a}$ parametrization where the
 evolution equations read: 
\begin{eqnarray}
&& \dot{\zeta} = - \frac{3 H}{2 \dot{H}} [ H (\delta\xi_{\rm a} + \delta\xi_{\rm b}) +(\dot{\overline{\xi}}_{\rm a} +\dot{\overline{\xi}}_{\rm b})\zeta] 
- \frac{1}{2 \dot{H}}(w_{\rm b} - w_{\rm a}) ( H \rho_{\rm a} \delta_{\rm a} + 
\dot{\rho}_{\rm a} \zeta),
\label{zetaex2}\\
&&
\dot{\delta}_{\rm a} + \frac{9 H^2}{\rho_{\rm a}} \biggl( \overline{\xi}_{\rm a} \delta_{\rm a} - \delta \xi_{\rm a} + \overline{\xi}_{\rm a} \frac{\dot{H}}{H^2} \zeta \biggr)= - \overline{\Gamma} (1 + w_{\rm a}) 
\biggl( \delta_{\Gamma} + \frac{\dot{H}}{H^2} \zeta \biggr).
\label{deltaex1}
\end{eqnarray}

Notice that Eqs. (\ref{zetaex1}) and (\ref{zetaaex1}) can be swiftly deduced from 
Eqs. (\ref{zetagi}) and (\ref{zetaagi}), by expressing the gauge-invariant 
fluctuations directly in the off-diagonal gauge. The only non-trivial 
expressions are 
\begin{equation}
{\cal H} \Xi + \overline{\xi}' \Psi = {\cal H} \delta \xi,
\label{exex1}
\end{equation}
and 
\begin{equation}
{\cal H} \Xi_{\rm a} + \overline{\xi}'_{\rm a} \Psi = {\cal H} \delta \xi_{\rm a}.
\label{exex2}
\end{equation}
The meaning of Eqs. (\ref{exex1}) and (\ref{exex2}) is that the two (gauge-invariant)
quantities at the left-hand sides translate, in the off-diagonal gauge, 
simply into the spatial variations of the bulk viscosity coefficient.
The right-hand sides of these equations are 
deduced from the left-hand 
sides by simply using Eqs. (\ref{PSI})--(\ref{XI}) (or Eqs. (\ref{PSI})--(\ref{XIab}))
and by setting, in the obtained expression,
$E=0$ and $\psi=0$.

For some applications,  it is 
interesting to consider situations where $\dot{H}\to 0$, at least 
for an interval of cosmic time. This 
may happen if the background solutions have, in some limit, a quasi-de 
Sitter evolution where $\dot{H}^{-1}$ may become very large. In this case 
the evolution equations in terms of $\zeta$ is partially invalidated. The evolution equations in terms of $\phi$ 
are, instead, fully valid and well defined in all their limits.

From Eqs. (\ref{OD3}), (\ref{OD4}) and (\ref{OD5}), neglecting the spatial gradients 
and using the cosmic time coordinate the relevant equations become
\begin{eqnarray}
&& \dot{\phi} + \biggl( 3 H + 2 \frac{\dot{H}}{H}\biggr)\phi = 
\frac{1}{2 H} \{ \delta p +
3 H [( \overline{\xi}_{\rm a} + \overline{\xi}_{\rm b})\phi 
- (\delta \xi_{\rm a} + \delta\xi_{\rm b})] \},
\nonumber\\
&& \dot{\delta}_{\rm a} + \frac{9 H^2}{\rho_{\rm a}} [ \overline{\xi}_{\rm a} \delta_{\rm a} - 
\delta \xi_{\rm a} + \overline{\xi}_{\rm a} \phi ] = - \overline{\Gamma} (w_{\rm a } + 1) [ \delta_{\Gamma} + \phi] ,
\nonumber\\
&& \dot{\beta} + H\beta  + \phi=0,
\end{eqnarray}
where $ a B = \beta$.

\renewcommand{\theequation}{6.\arabic{equation}}
\section{Physical applications}
\setcounter{equation}{0}
\subsection{The case of vanishing background viscosity}
Two complementary situations may arise. 
In the first case the homogeneous 
component of the bulk viscosity coefficient vanishes  
(i.e. $ \overline{\xi}_{\rm a} 
=\overline{\xi}_{\rm b} =0$) but the inhomogeneous part does not 
vanish (i.e. either $\delta \xi_{\rm a} \neq 0$ or $\delta\xi_{\rm b} \neq 0$).
In this case it is plausible, from either Eqs. (\ref{zetagi}), (\ref{zetaagi}) or (\ref{exex1}), (\ref{exex2}), that the evolution of curvature perturbations 
will be supplemented by a source term proportional 
to the spatial variations of the bulk viscosity coefficient.

This situation is, to some extent, 
 a bit arbitrary, but it closely analogous to the situation encountered in the case of a decay rate with spatial 
fluctuations, i.e.  the toy model when the decay rate is allowed to be 
inhomogeneous but without a specific microscopic 
model supporting this interpretation \cite{dvali}. In spite of this 
important caveat let us investigate first this (rather extreme) situation. 

\begin{figure}
\begin{center}
\begin{tabular}{|c|c|}
      \hline
      \hbox{\epsfxsize = 7 cm  \epsffile{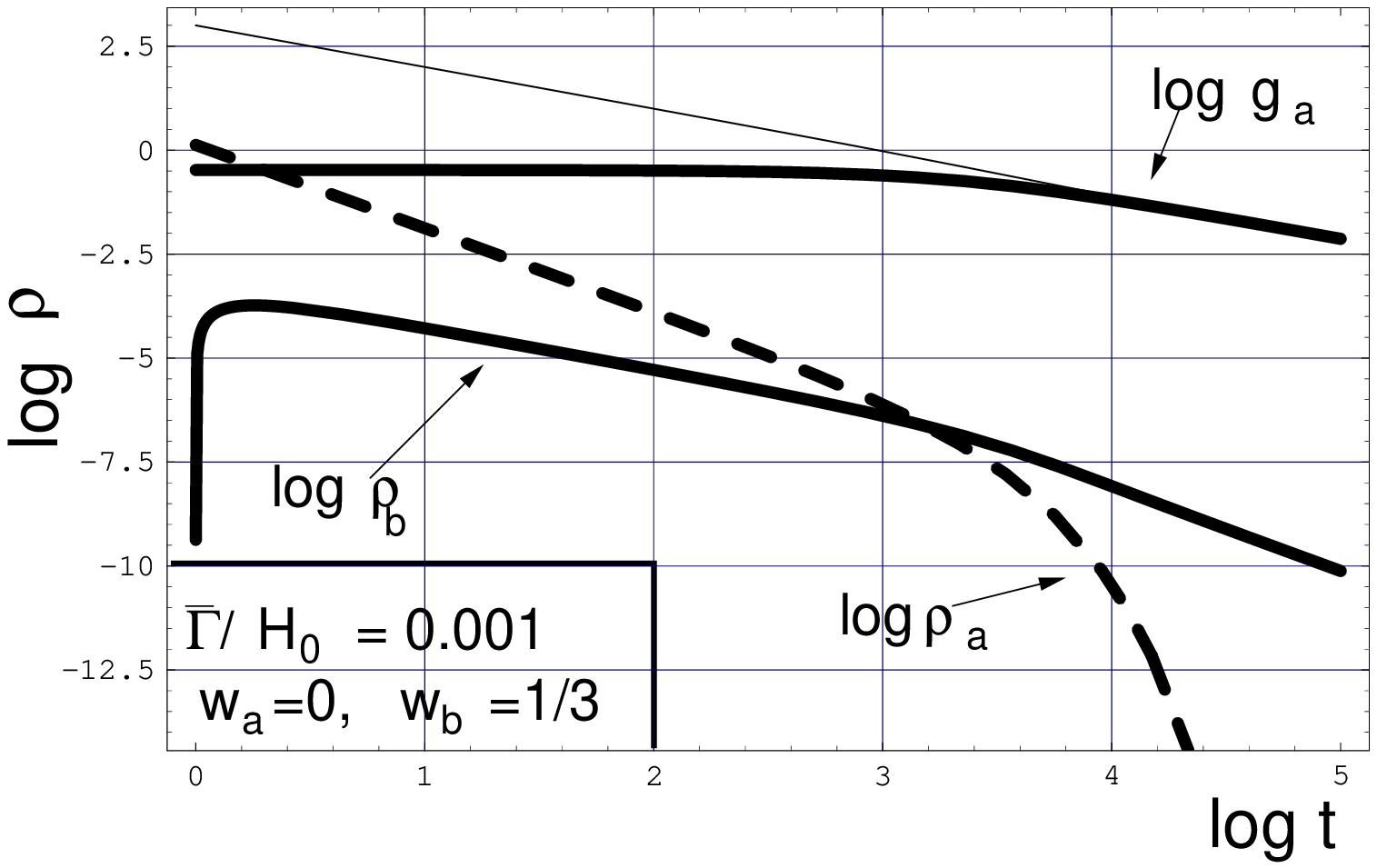}} &
      \hbox{\epsfxsize = 7 cm  \epsffile{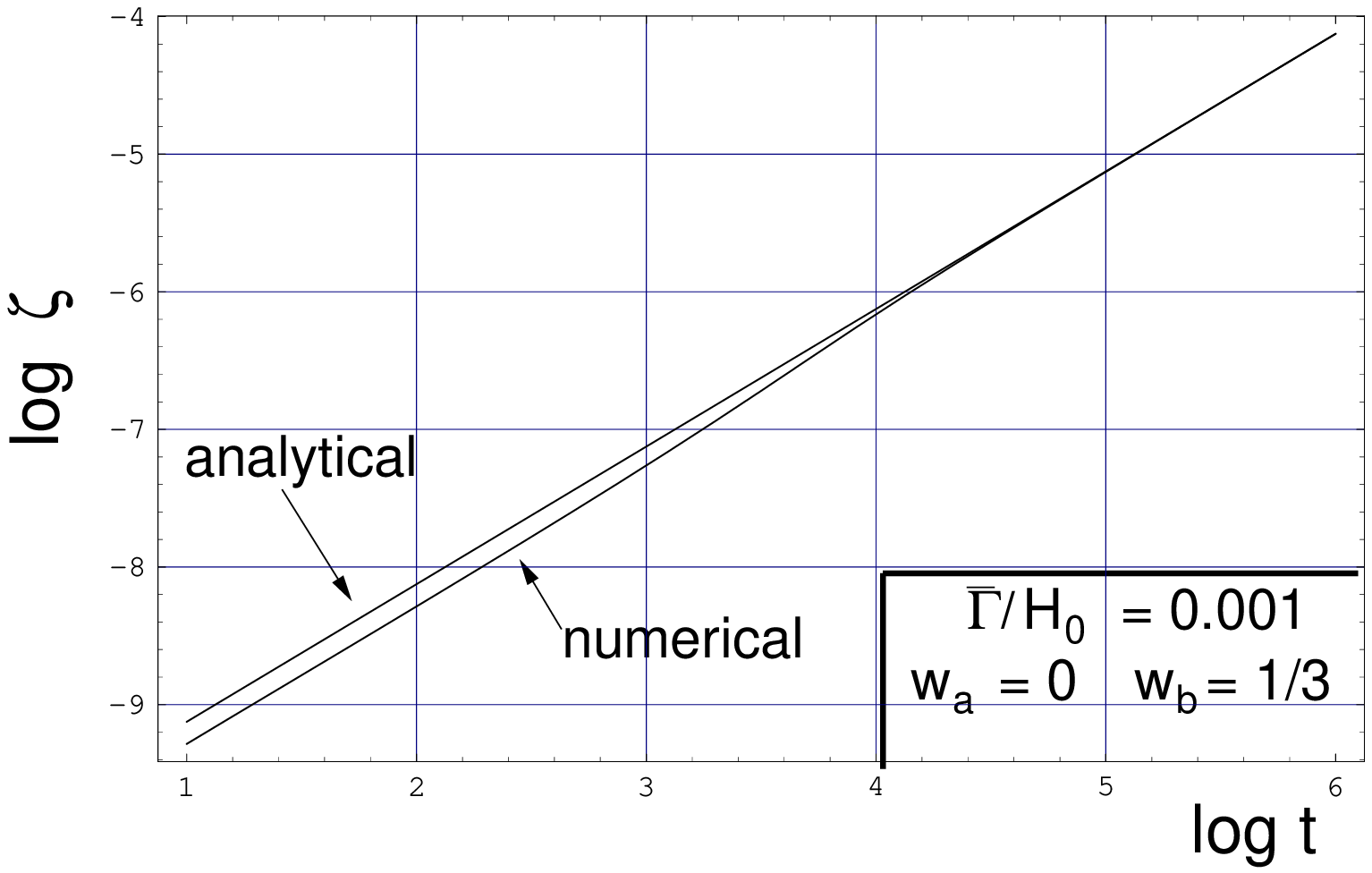}}\\
      \hline
\end{tabular}
\end{center}
\caption[a]{The evolution of the background (left-hand-side plot) and of the fluctuations (right-hand-side plot)
are illustrated in various cases of physical interest. 
The parameters are fixed in so that $\Gamma/H_{0} = 10^{-3}$, $\delta_{\Gamma}= 10^{-7}$, 
$\delta \xi_{\rm b} = 10^{-10}$, $\delta\xi_{\rm a} =0$. In this figure (and in the following) $H_{0}$ denotes the value of the Hubble parameter at the initial 
integration time. }
\label{F1}
\end{figure}

Suppose, for simplicity, that $\overline{\xi}_{\rm a}= \overline{\xi}_{\rm b} =0$ 
but $\delta \xi_{\rm b} \neq 0$. In this case, it can be shown that, in spite of the 
value of $\delta_{\Gamma}$, $\zeta$ increases linearly in time. In Fig. \ref{F1} 
the integration of Eqs. (\ref{zetaex1}) and (\ref{zetaaex1}) is illustrated. 
In the  plot at the left-hand side, the evolution of the energy densities of the background is reported
for the case where the decay products are given by a radiation fluid (i.e. $w_{\rm b} = 1/3$) and the a-fluid is given by pressureless matter ($w_{\rm a} =0$).
Similar plots can be obtained when $w_{\rm a} \neq 0$ (for instance 
$w_{\rm a} =1$ or $w_{\rm a} =1/10$). In the present context 
it will often be interesting to fix $w_{\rm b}$ to $1/3$ and to leave 
$w_{\rm a}$ free to vary form $0$ to $1$. 
As  is clear from Fig. \ref{F1}, for $ t > \overline{\Gamma}^{-1}$, 
the energy density of the a-fluid is exponentially 
damped thanks to the finite value of the decay. For curvature scales 
much smaller than the decay rate the radiation background dominates. 
In Fig. \ref{F1} (plot at the right-hand side) the evolution 
of $\zeta$ is compared with the analytical approximation valid for large times.
 
To get an analytical approximation for the evolution of $\zeta$ let us start 
from Eqs. (\ref{zetaex1}) and (\ref{zetaaex1}). We are interested in the 
case where $\overline{\xi}_{\rm a} = \overline{\xi}_{\rm b} =0$. Moreover, as 
assumed above, we may also posit, for simplicity,  $w_{\rm a} =0$ (which is the case numerically treated 
in Fig. \ref{F1}). Under these assumptions, Eqs. (\ref{zetaex1}) and (\ref{zetaaex1})
become, respectively 
\begin{eqnarray}
&&\dot{\zeta} = \frac{w_{\rm b}}{2 \dot{H}} \dot{\rho}_{\rm a} (\zeta_{\rm a} - \zeta)  - \frac{3 H^2}{2 \dot{H}} \delta\xi,
\label{detail1}\\
&& \dot{\zeta}_{\rm a} = \overline{\Gamma} \frac{\dot{H}}{H^2} g_{\rm a} ( \zeta_{\rm a} - \zeta) - \overline{\Gamma} g_{\rm a} \delta_{\Gamma}.
\label{detail2}
\end{eqnarray}
To proceed further, notice that, from Eq. (\ref{defga}) we have, in the 
case $w_{\rm a} =0$ 
\begin{equation}
g_{\rm a} = - \frac{H \rho_{\rm a}}{\dot{\rho}_{\rm a}} = 
\frac{H}{3 H + \overline{\Gamma}}
\label{detail3}
\end{equation}
where the second equality follows from the first by explicit 
use of the (background) continuity equation for the a-fluid 
 in the case $w_{\rm a}=0$:
\begin{equation}
\dot{\rho}_{\rm a} + ( 3 H + \overline{\Gamma}) \rho_{\rm a} =0. 
\label{detail4}
\end{equation}
Since $\overline{\Gamma}$ is constant,  from 
Eq. (\ref{detail3}) it is clear that, for large times,
\begin{equation}
g_{\rm a} \simeq H/\overline{\Gamma}. 
\label{detail4a}
\end{equation}
This 
asymptotic expression compares very well with the 
numerical result. In Fig. \ref{F1} (plot at the left), the first thick 
line from the top corresponds to the numerical evolution 
of $\log{g_{\rm a}}$ while the (first) thin line (from the top) 
is the result obtained by approximating 
$g_{\rm a} \sim H/\overline{\Gamma}$.  

Notice, furthermore, that for sufficiently large times (i.e. $t > \overline{\Gamma}^{-1}$) the evolution of $H$ is simply given by 
\begin{equation}
H \simeq \frac{2}{3(w_{\rm b} + 1) t}. 
\label{detail5}
\end{equation}
This is because the contribution of the a-fluid to the Hubble rate 
vanishes exponentially for $t > \overline{\Gamma}^{-1}$. In fact, from 
Eq. (\ref{detail4}) direct integration leads to 
\begin{equation}
\rho_{\rm a}(t) = \rho_{\rm a}(t_{0}) \biggl(\frac{a_0}{a}\biggr)^3 e^{ - \overline{\Gamma} (t -t_{0})}.
\label{detail6}
\end{equation}
Equations (\ref{detail4})--(\ref{detail4a}) and (\ref{detail5})--(\ref{detail6}) 
allow to estimate analytically the time-dependent coefficients appearing 
in eqs. (\ref{detail1}) and (\ref{detail2}).
In particular, by looking at Eq. (\ref{detail1}), the coefficient of the 
first term at the right hand side 
is exponentially suppressed, for $t\ > \overline{\Gamma}^{-1}$ 
since it contains $\dot{\rho}_{\rm a}$ (see Eq. (\ref{detail6})).
On the contrary, the coefficient of the second term at the right hand side 
of Eq. (\ref{detail1}) is finite and it is given by (see Eq. (\ref{detail5})) 
\begin{equation}
- \frac{3}{2} \frac{H^2}{\dot{H}} = \frac{1}{w_{\rm b} + 1}. 
\label{deltail7}
\end{equation}
Since $\delta \xi$ is constant in time, direct integration of Eq. (\ref{detail1}) 
leads to
\begin{equation}
\zeta \simeq \frac{\delta \xi}{w_{\rm b } + 1} \biggl(\frac{t}{t_{1}}\biggr).
\label{detail8}
\end{equation}
where $t_1 \geq \overline{\Gamma}^{-1}$. 
By keeping the dominant 
terms in the limit $t \gg \overline{\Gamma}^{-1}$ Eq. (\ref{detail2}) 
allows also to determine the evolution of $\zeta_{\rm a}$ that turns out to be 
of the same order of $\zeta$.
Equation (\ref{detail8}) is illustrated in the right-hand plot of Fig. \ref{F1}  and it compares very well with the numerical solution. 

\subsection{The case of non vanishing background viscosity}

The following example we ought to discuss is the one where 
$\xi_{\rm a} = \epsilon \sqrt{\rho_{\rm a}}$ and $\xi_{\rm b} =0$. 
In this case, 
\begin{equation}
\delta \xi_{\rm a} = \epsilon\frac{\delta \rho_{\rm a}}{2 \sqrt{\rho_{\rm a}}},\,\,\,\,\,\,\,\,\,
\dot{\overline{\xi}}_{\rm a} = \epsilon\frac{ \dot{\rho}_{\rm a}}{2\sqrt{\rho_{\rm a}}}.
\end{equation}
Moreover, in the off-diagonal gauge, the following chain of equality 
holds 
\begin{equation}
\delta \xi_{\rm a} =\epsilon  \frac{\sqrt{\rho_{\rm a}}}{2} \delta_{\rm a} = - \epsilon
\frac{\dot{\rho_{\rm a}}}{ 2 H \sqrt{\rho_{\rm a}}} \zeta_{\rm a}.
\end{equation}
It will be relevant, to physical applications, to analyse the case 
when the a-fluid decays into radiation (i.e. $w_{\rm b} = 1/3$) 
with different values of the homogeneous decay rate and different 
initial values of $w_{\rm a}$.  When the a-fluid decays, for $\Gamma \sim H$,
energy is transferred to the b-fluid. Different 
values of $\epsilon$ affect the overall normalization of the energy density 
of the a-fluid. However, since the a-fluid decays exponentially and the 
b-fluid decays as $a^{- 3 (w_{\rm b} +1)}$ (where $a$ is the scale factor) 
the a-fluid will quickly be subdominant. Different values of the 
initial decay rate (i.e. $\overline{\Gamma}/H_{0} < 1$) may shift 
the onset of the b-dominated phase. It is relevant to stress, at this point, 
that initial conditions for the evolution of the background are set 
in such a way that $\rho_{\rm b}(t_{0}) =0$ (where $t_{0}$ is the 
initial integration time). Initially, around $t_{0}$, the a-fluid dominates 
and it can easily be shown, recalling that 
$\overline{\xi}_{\rm a} = \epsilon \sqrt{\rho_{\rm a}}$ that the approximate 
solution for the evolution of the background, in this regime, is given by 
\begin{equation}
H(t) \simeq \frac{2}{[3 (w_{\rm a} + 1) - 3 \sqrt{3} \epsilon]t},\,\,\,\,\,\,\,\,\,\,\,
\rho_{\rm a}(t) \sim 3 H^2(t),\,\,\,\,\,\,\,\, t_{0} \leq t < \overline{\Gamma}^{-1}.
\label{detail1a}
\end{equation}
 In fact, for  $t < \overline{\Gamma}^{-1}$ 
the a-fluid dominates the energy-momentum tensor, therefore from Eqs. (\ref{B1})--(\ref{B2}) and (\ref{Ac1})--(\ref{Bc1}) (written in units $8\pi G =1$ that 
are the units employed in our numerical codes) we can obtain 
the following equation 
\begin{equation}
\dot{H} = - \frac{3}{2} [ (w_{\rm a} + 1) - \sqrt{3} \epsilon] H^2 
\label{detail2a}
\end{equation}
whose integral with respect to cosmic time is exactly Eq. (\ref{detail1a}).
For $  t_{0} \,\laq\, t < \overline{\Gamma}^{-1}$, 
the scale factor $a(t)$  expands  (i.e. $\dot{a} >0$) in a 
decelerated way (i.e. $\ddot{a} <0$) and $\dot{H}$ never goes 
to zero iff  $\epsilon < (w_{\rm a} + 1)/\sqrt{3}$.  Notice that, in the limit 
$\epsilon \to 0$ Eq. (\ref{detail1a}) is simply the expansion rate of a 
Universe filled with a perfect fluid $\rho_{\rm a}$. We stress that this solution 
is only valid in the neighborhood of the initial integration time $t_{0}$: later on 
the full numerical solution is mandatory.
\begin{figure}
\begin{center}
\begin{tabular}{|c|c|}
      \hline
      \hbox{\epsfxsize = 7 cm  \epsffile{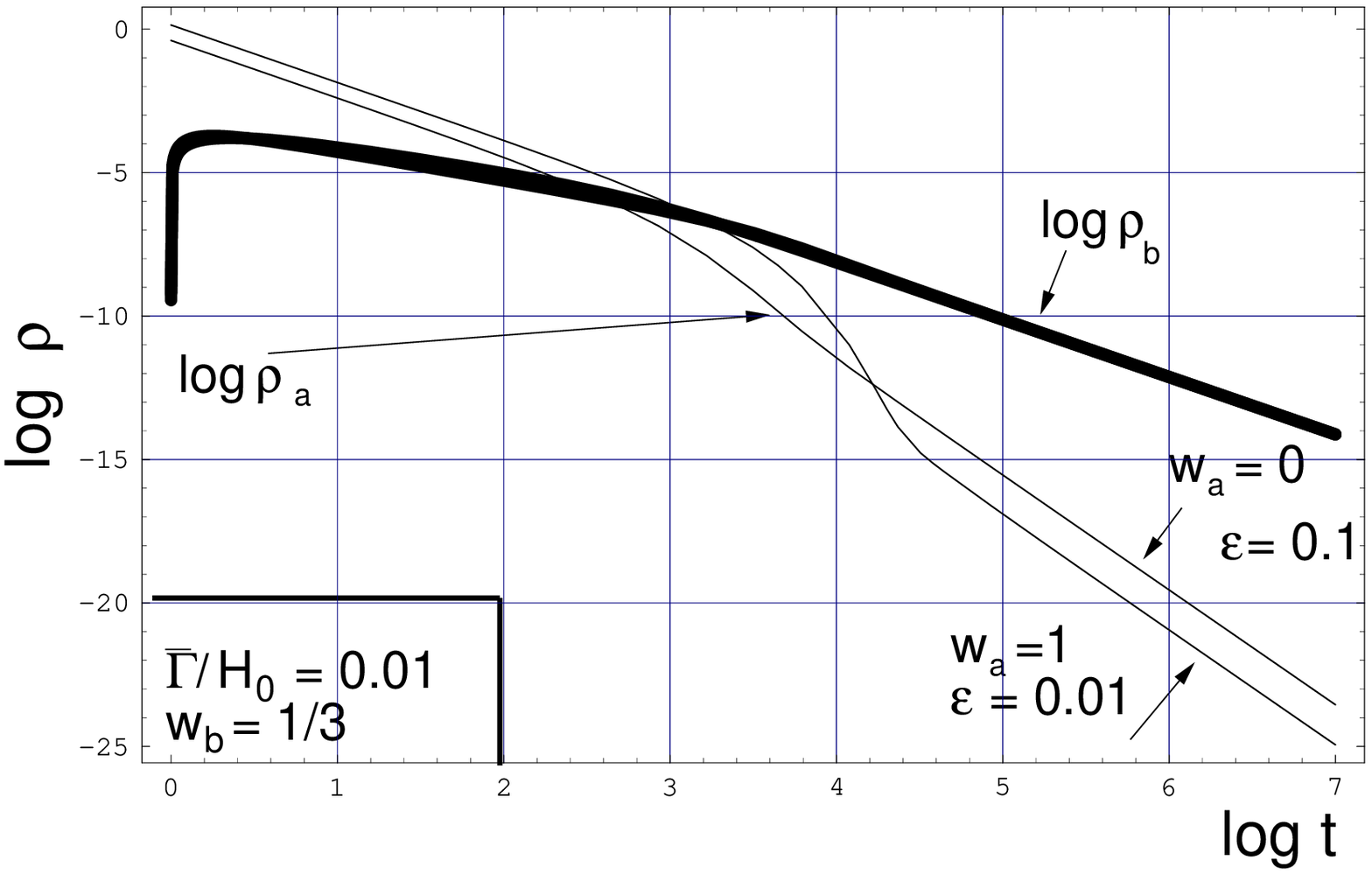}} &
      \hbox{\epsfxsize = 7 cm  \epsffile{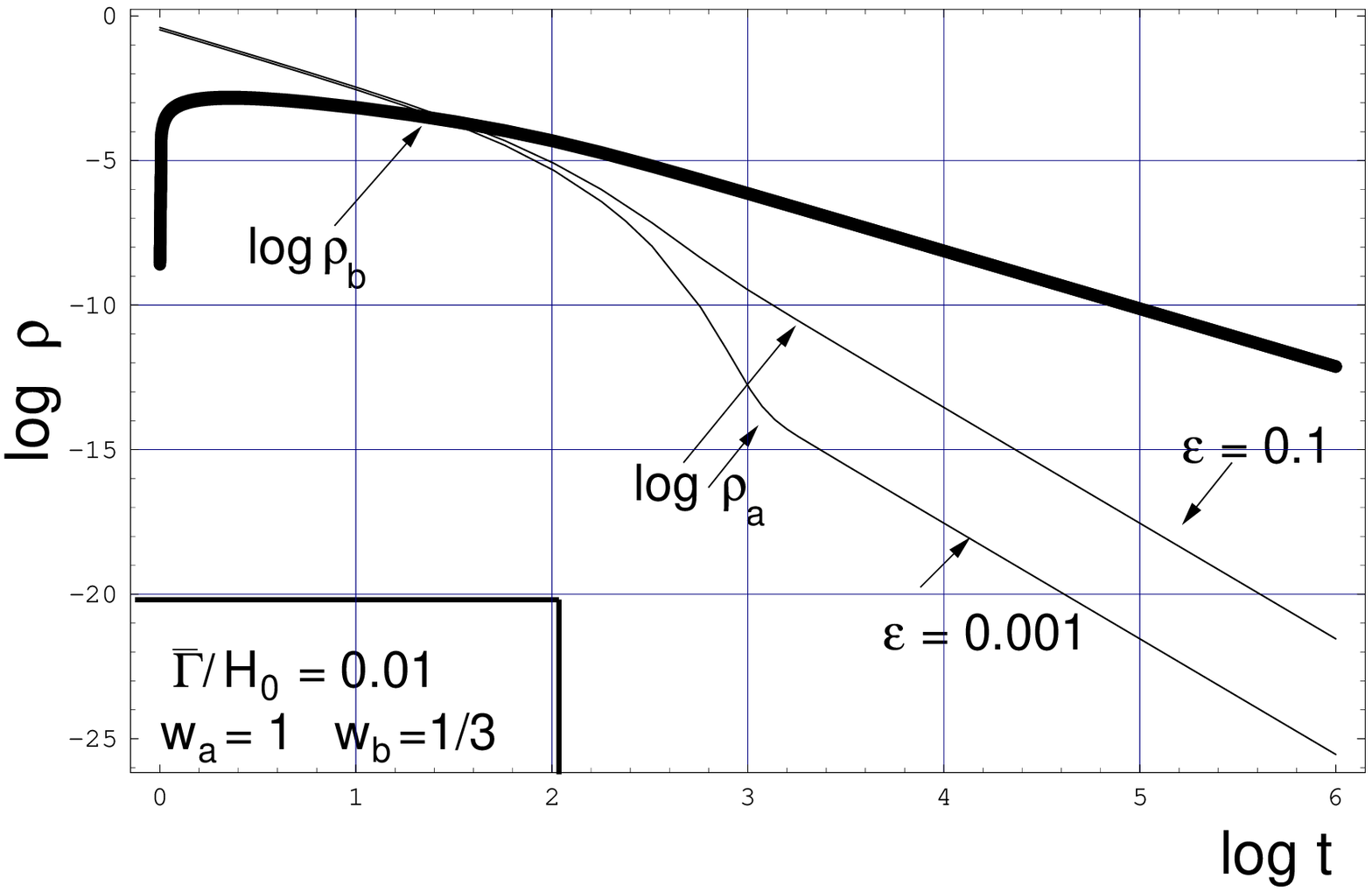}}\\
      \hline
\end{tabular}
\end{center}
\caption[a]{Evolution of the energy densities of the a- and 
b-fluid for different choices of the parameters in the case when 
the homogeneous bulk viscosity coefficient is parametrized as 
$\overline{\xi}_{\rm a} = \epsilon \sqrt{\rho_{\rm a}}$. The thin lines 
always indicate the common logarithm of the energy 
density of the a-fluid, while the thick lines refer to the common logarithm 
 of the energy density of the b-fluid (that 
dominates for $t \gg \overline{\Gamma}^{-1}$).}
\label{F2}
\end{figure}
The mentioned features of the background solutions are 
illustrated in Fig. \ref{F2}, where the evolution of the 
common logarithms of the energy density are 
reported as a function of the common logarithm of 
the cosmic time coordinate \footnote{In the remaining part 
of the article the logarithms will always be 
common logarithms, i.e. logarithms to base 10.}. 
As a side remark, we may note that in Fig. \ref{F2} (plot on
 the left-hand side)
the first two lines at the top illustrate the behaviour of the 
function $g_{\rm a}(t)$ (defined in Eq. (\ref{defga})) and of 
its approximation (valid for large times) i.e. $H/\overline{\Gamma}$.
The function $g_{\rm a}$ is always regular  and never vanishes. 
This aspect is related to the occurrence that the contribution 
of the bulk viscosity becomes  negligible, in practice,  when the b-fluid 
dominates.

\begin{figure}
\begin{center}
\begin{tabular}{|c|c|}
      \hline
      \hbox{\epsfxsize = 7 cm  \epsffile{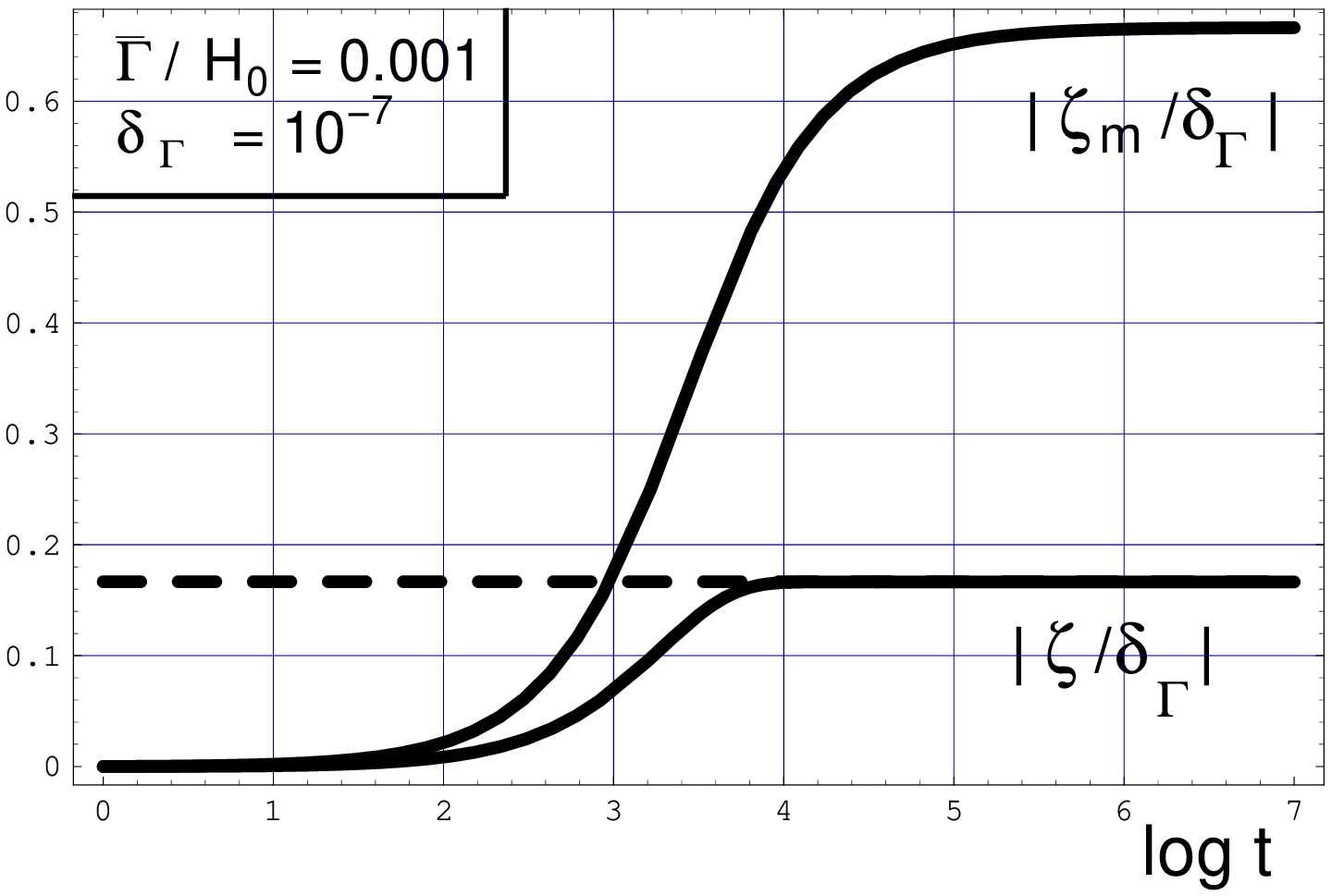}} &
      \hbox{\epsfxsize = 7 cm  \epsffile{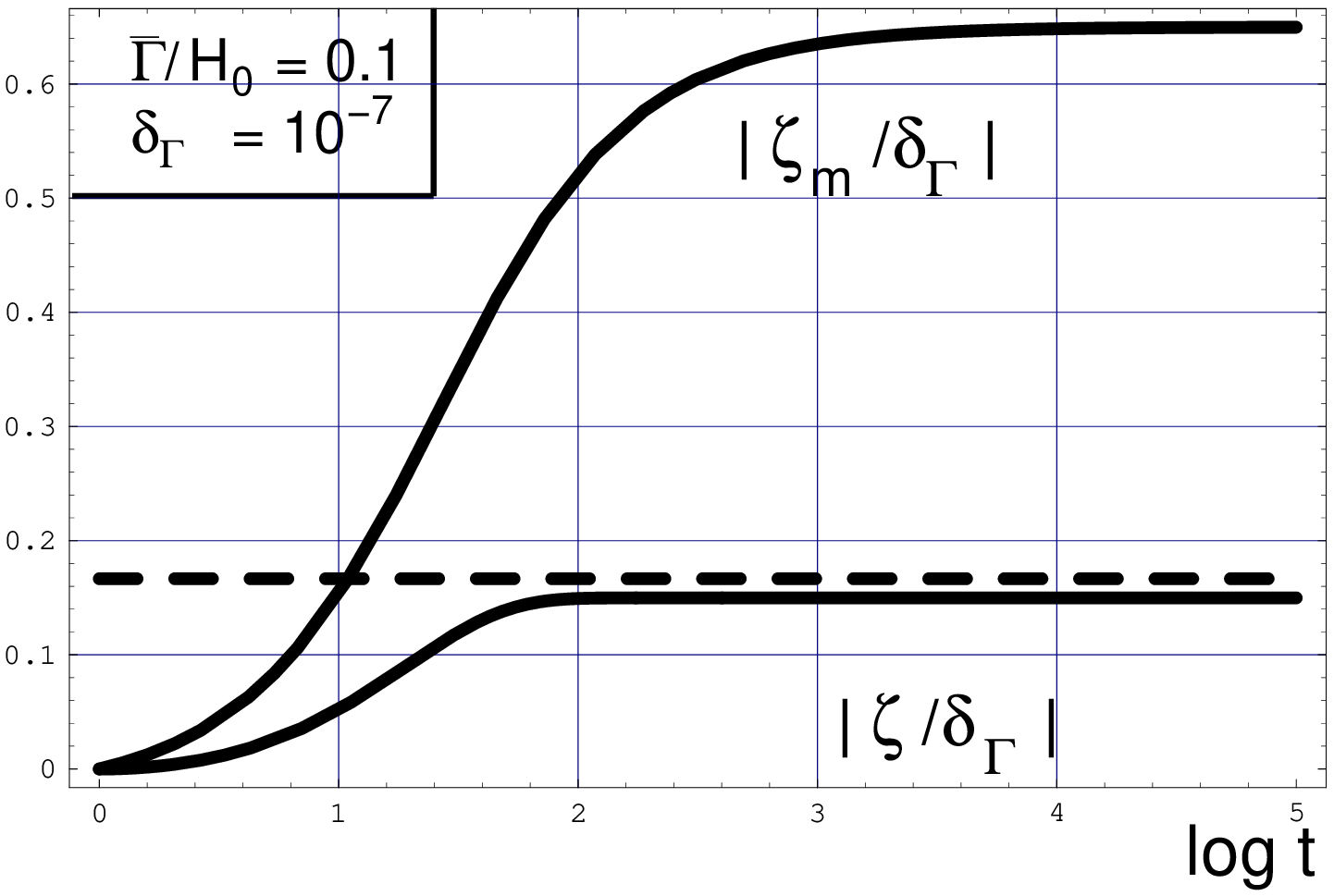}}\\
      \hline
\end{tabular}
\end{center}
\caption[a]{Evolution of curvature fluctuations in the case $\epsilon =0$ 
and for different values of the homogeneous component 
of the decay rate.}
\label{F3}
\end{figure}

It is now appropriate to analyse the evolution of $\zeta$ and $\zeta_{\rm a}$.
Consider, to begin with, the case where $\epsilon =0$.  Following 
the discussion of \cite{dvali} (see also \cite{mazumdar}), the asymptotic value 
of $\zeta$ is $\zeta_{\rm final} \simeq - \delta_{\Gamma}/6$. This is exactly what is 
found numerically (see Fig. \ref{F3}: lower curves in both plots 
at the left- and right-hand sides). In Fig. \ref{F3} the predicted asymptotic value 
is reported with a dashed line. As already remarked in \cite{dvali} 
the accuracy of the estimate is controlled by the ratio $\overline{\Gamma}/H_{0}$.
This aspect is also illustrated in Fig. \ref{F3}. In the plot on
 the left-hand side 
$\Gamma/H_{0} = 10^{-3}$, while on the right-hand side 
$ \Gamma/H_{0} = 0.1$. Even if the asymptotic value of
 $|\zeta/\delta_{\Gamma}|$ is always 
well approximated by $1/6$, the prediction is, comparatively, less accurate when $\Gamma/H_{0}$ is larger.  

When $\epsilon \neq 0$ and 
the discussion of \cite{dvali}  can be 
generalized.  The authors of Ref. \cite{dvali} notice that if $\overline{\Psi}$ 
is the solution of the system for $\delta_{\Gamma} =0$, then the 
``forced" solution (corresponding to the case $ \delta_{\Gamma}\neq 0$ ) can 
be obtained as $ \Psi = \overline{\Psi} - \delta_{\Gamma}$.
Assuming that, indeed, for large times both the curvature 
fluctuations $\zeta$ and the Bardeen potential $\Psi$ 
 go to a constant (as it is the case from the numerical 
analysis), we have that the final asymptotic value of $\Psi$ in the  case 
where $\delta_{\Gamma} \neq 0$ and $\epsilon \neq 0$ is given 
by 
\begin{equation}
\Psi_{\rm final} = \frac{[ ( 3 w_{\rm a} + 5 ) - 3 \sqrt{3} \epsilon] (w_{\rm b} + 1) 
- ( 3 w_{\rm b} + 5) [ (w_{\rm a} + 1) - \sqrt{3} \epsilon]}{ ( 3 w_{\rm b} + 5)[ (w_{\rm a} + 1) - \sqrt{3} \epsilon]} \,\, \delta_{\Gamma}.
\label{detail2e}
\end{equation}
Recall now that, when the a-fluid has already decayed, the relation 
between $\zeta$ and $\Psi$ (up to spatial gradients, i.e. for length-scales 
larger than the Hubble radius) is nothing but 
\begin{equation}
\zeta_{\rm final} = - \frac{3 w_{\rm b} + 5}{3(w_{\rm b} + 1)} \Psi_{\rm final}.
\label{detail2f}
\end{equation} 
From Eqs. (\ref{detail2e}) and (\ref{detail2f}), the 
final value of curvature fluctuations is then given by the following 
general equation:
\begin{equation}
\zeta_{\rm final} = -  \biggl\{ \frac{(w_{\rm b} + 1)( 3 w_{\rm a} + 5 - 3 \sqrt{3} \epsilon)- 
(3 w_{\rm b} + 5) [(w_{\rm a} + 1) - \sqrt{3} \epsilon)]}{3(w_{\rm b} + 1) [ (w_{\rm a} + 1) - \sqrt{3} \epsilon]}\biggr\} \delta_{\Gamma},
\label{zetafinal}
\end{equation}
where by $\zeta_{\rm final}$ we mean, as before, the final asymptotic value 
in the phase dominated by the decay products, i.e. the b-fluid. 

In the case $w_{\rm a} = 0$ (leaving $w_{\rm b}$ undetermined)  Eq. 
(\ref{zetafinal}) leads to 
\begin{equation}
\zeta_{0\to w_{\rm b}} = - \frac{ [(w_{\rm b} +1) ( 5 - 3 \sqrt{3}\epsilon) - (3 w_{\rm b} + 5) ( 1 - \sqrt{3}\epsilon)}{3  (w_{\rm b} + 1) ( 1 - \sqrt{3}\epsilon)} \delta_{\Gamma}.
\label{zetafin2}
\end{equation}
In Eq. (\ref{zetafin2}) the subscript of $\zeta$ means that the quantity 
computed at the right-hand side is the asymptotic value of $\zeta$ in the case 
of the decay of $w_{\rm a} =0$ to $w_{\rm b} \neq w_{\rm a}$. The same 
notation will be used in the following equations.

In the case $w_{\rm b} = 1/3$ (leaving $w_{\rm a}$ undetermined) 
Eq. (\ref{zetafinal})
leads to 
\begin{equation}
\zeta_{w_{\rm a}\to 1/3} =  - \frac{1}{6} \biggl[\frac{ ( 1 - 3 w_{\rm a}) + 3 \sqrt{3} \epsilon}{w_{\rm a} + 1 - \sqrt{3} \epsilon}\biggr] \delta_{\Gamma}.
\label{zetafin3}
\end{equation}

By setting $w_{\rm b} =1/3$ in  Eq. (\ref{zetafin2}) we will get 
the result valid for the case of a mixture of dust matter and radiation, i.e. 
\begin{equation}
\zeta_{0\to 1/3} = - \frac{1}{6} \biggl(\frac{1 + 3 \sqrt{3} \epsilon}{1 - \sqrt{3} \epsilon}\biggr) 
\delta_{\Gamma}.
\label{zetafin4}
\end{equation}
If we set $w_{\rm a} = 1$ in Eq. (\ref{zetafin3}), we deduce that 
\begin{equation}
\zeta_{ 1\to 1/3} = \frac{1}{6} \frac{ 2 - 3 \sqrt{3} \epsilon}{2 - \sqrt{3} \epsilon} \delta_{\Gamma}.
\label{zetafin5}
\end{equation}

By setting $\epsilon \to 0$ in Eq. (\ref{zetafin4}) the usual result 
(valid in the absence of bulk viscosity) is reproduced.  It is an amusing
numerical coincidence that, 
in the limit $\epsilon \to 0$, $\zeta_{0\to 1/3} = - \zeta_{1\to 1/3}$, i.e. 
 $|\zeta_{0\to 1/3} |= | \zeta_{1\to 1/3}| =1/6$.

\begin{figure}
\begin{center}
\begin{tabular}{|c|c|}
      \hline
      \hbox{\epsfxsize = 7 cm  \epsffile{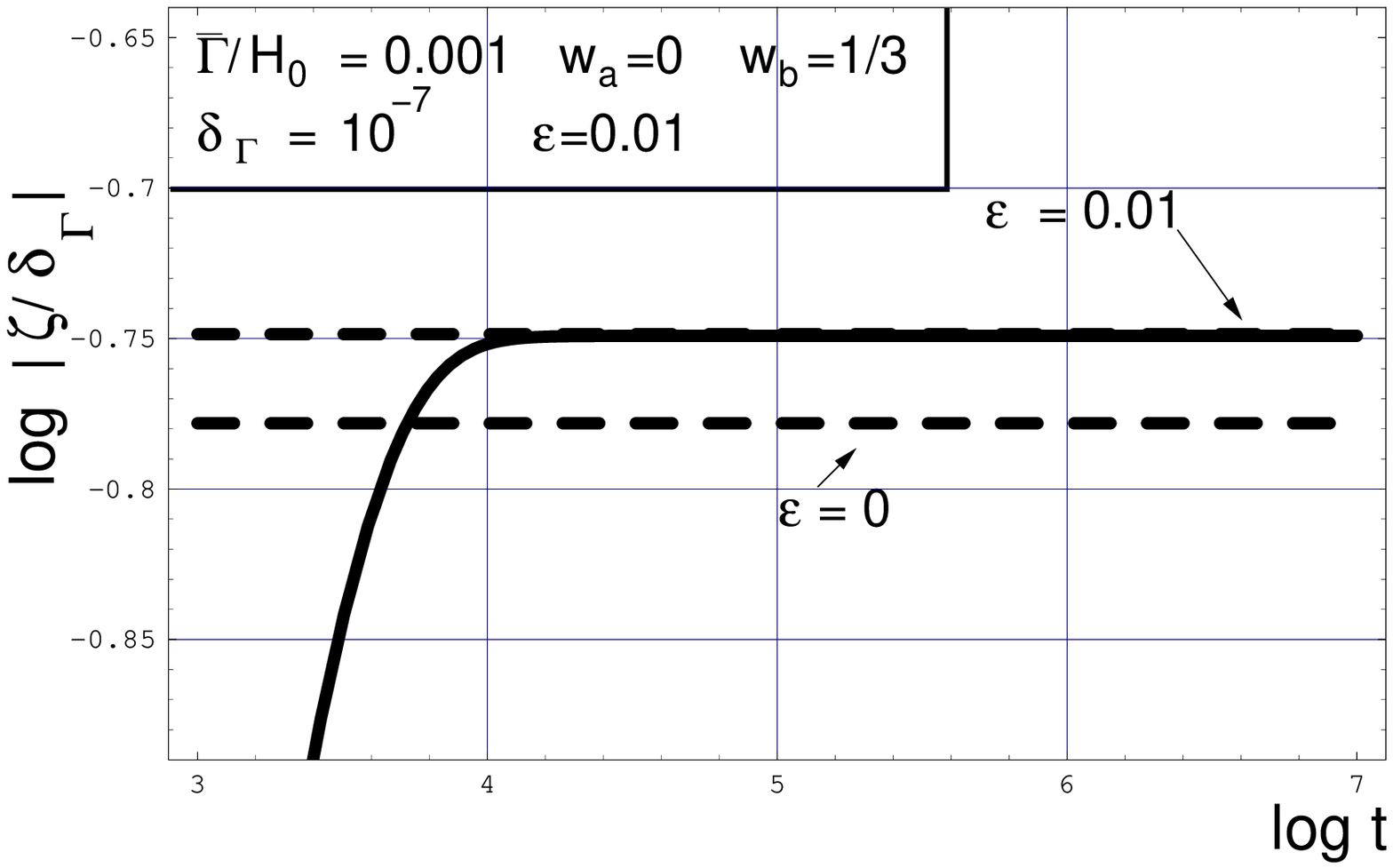}} &
      \hbox{\epsfxsize = 7 cm  \epsffile{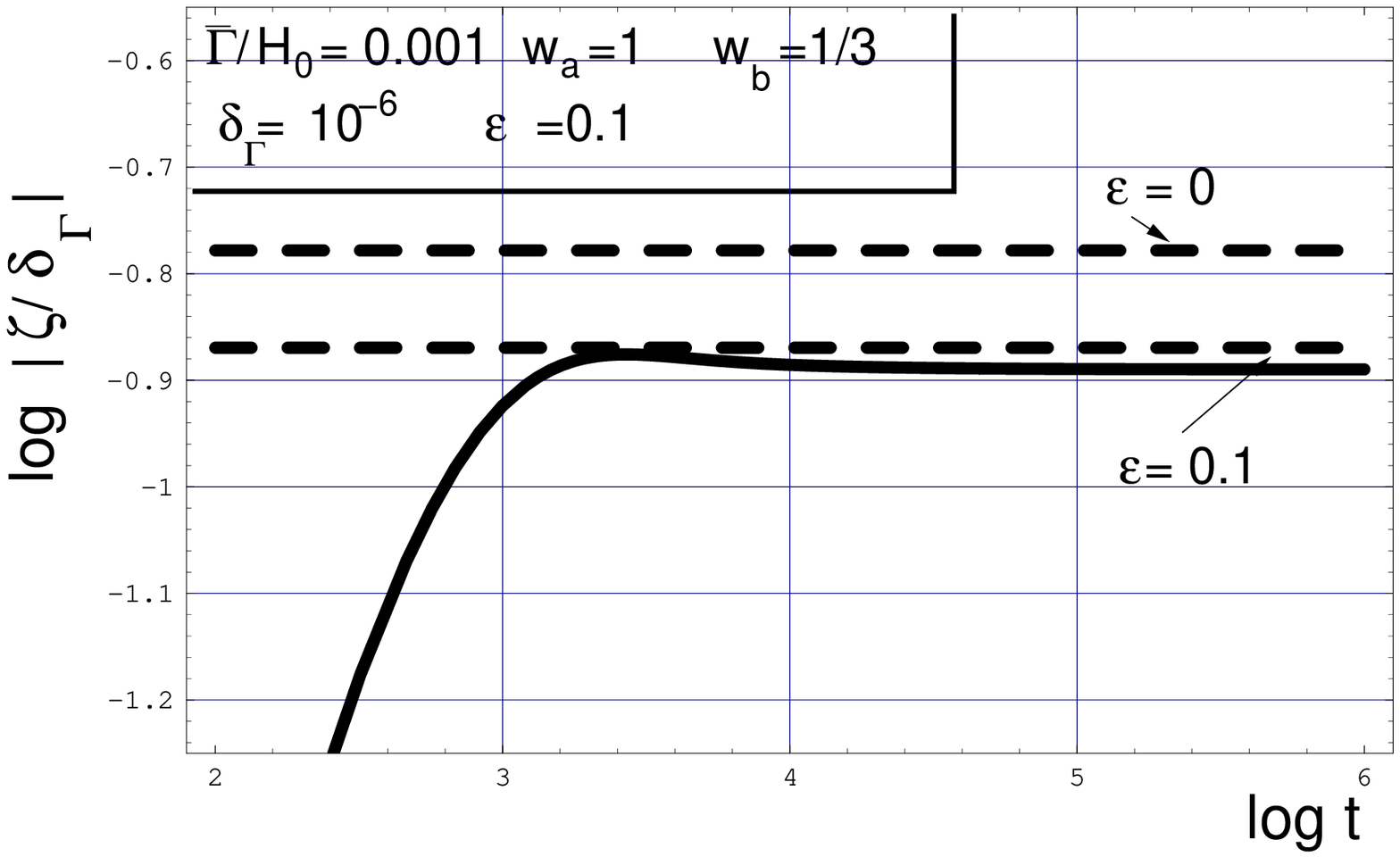}}\\
      \hline
\end{tabular}
\end{center}
\caption[a]{The analytical results obtained from Eq. (\ref{zetafinal}) are compared with 
the numerical calculation for the evolution of $\zeta$ for the transition 
dust-radiation (left plot) and for the transition stiff-radiation.}
\label{F4}
\end{figure}

\begin{figure}
\begin{center}
\begin{tabular}{|c|c|}
      \hline
      \hbox{\epsfxsize = 7 cm  \epsffile{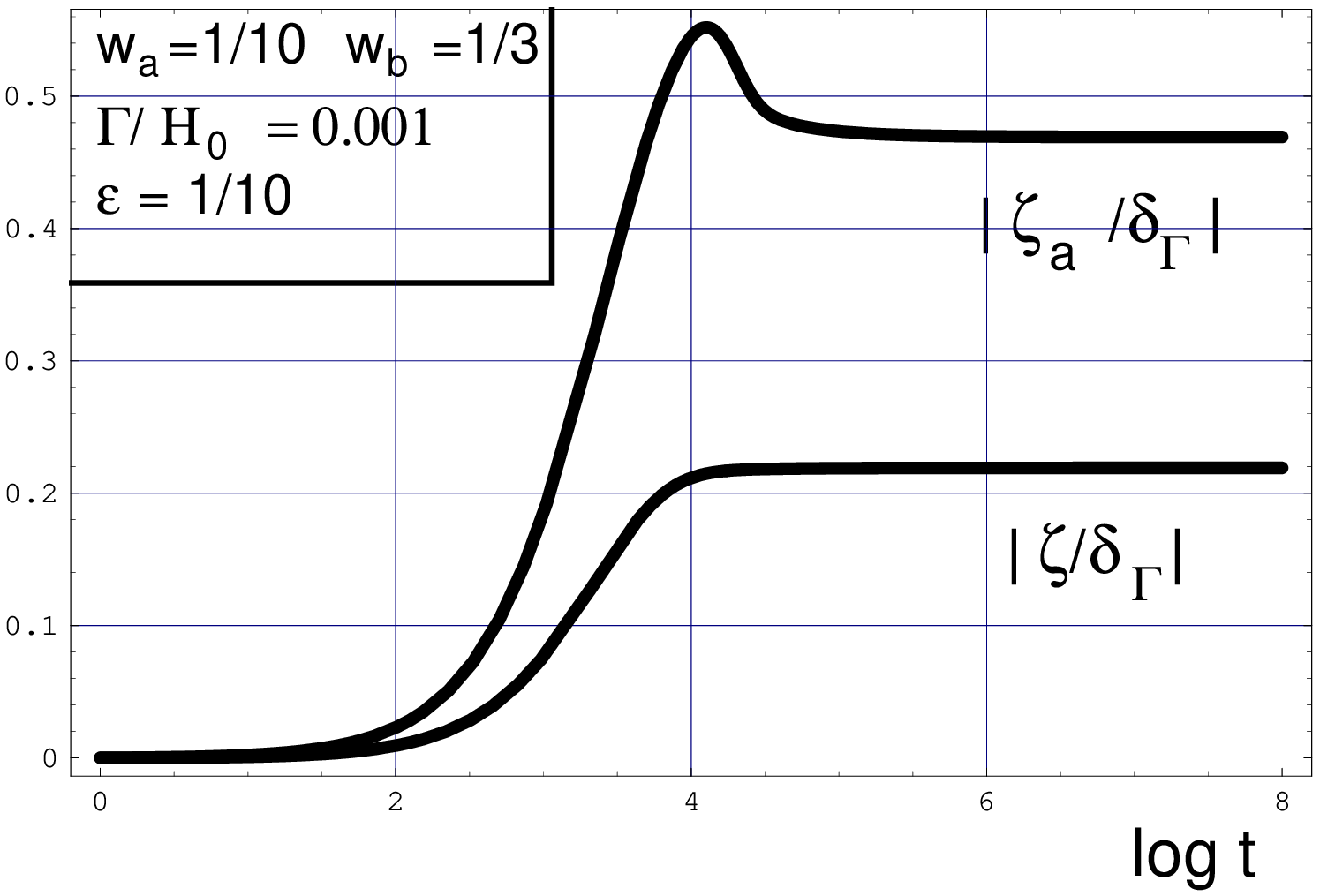}} &
      \hbox{\epsfxsize = 7 cm  \epsffile{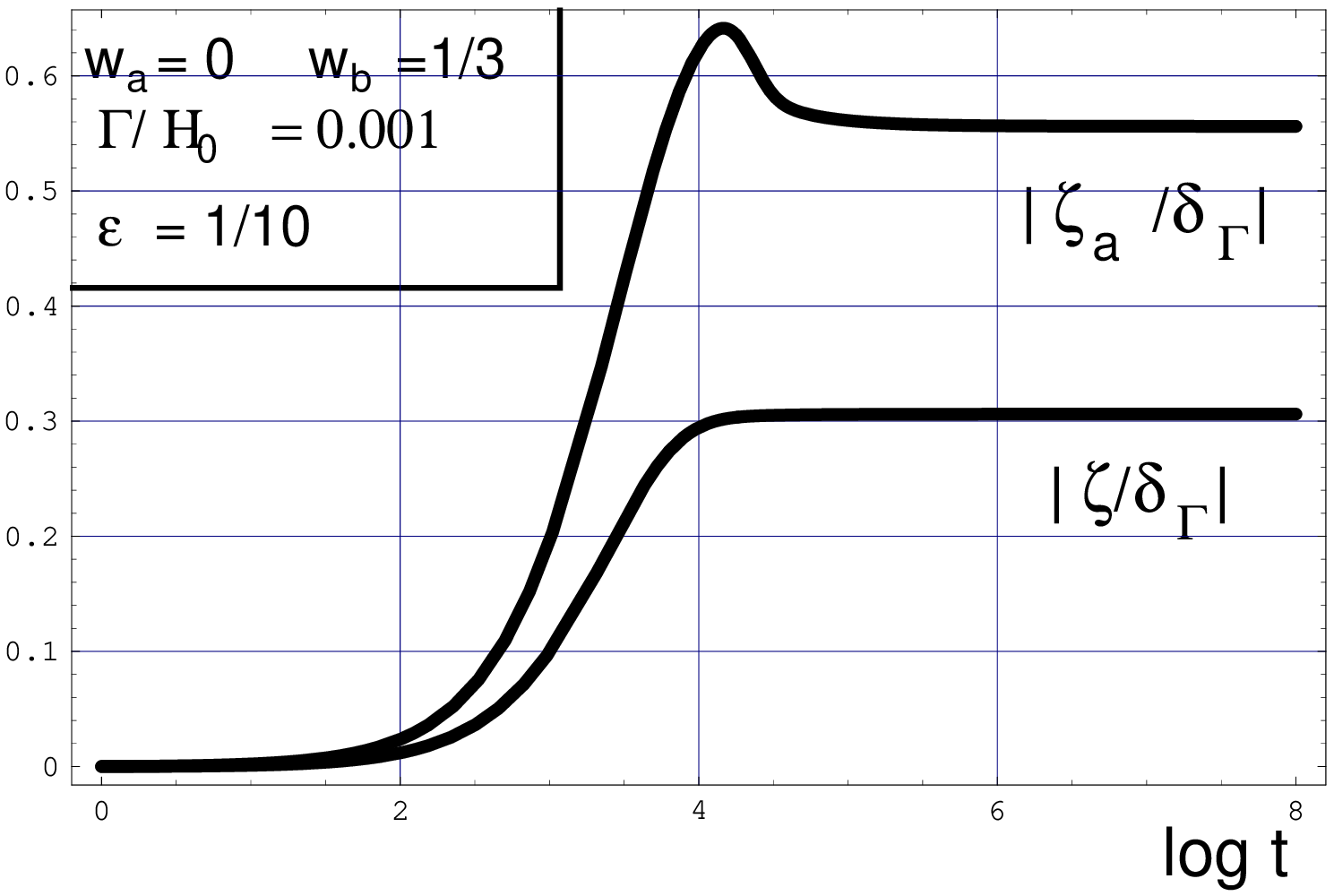}}\\
      \hline
\end{tabular}
\end{center}
\caption[a]{The integration of the evolution equations of the fluctuations is reported 
for different mixtures of fluids. In the  plot at the left $\delta_{\Gamma}=10^{-6}$,at the right $\delta_{\Gamma}=10^{-8}$.}
\label{F5}
\end{figure}

These results compare rather well with the numerical integration of the 
full system.  In Fig. \ref{F4} we report the result for the numerical integration of 
the fluctuations in terms of the logarithm of $|\zeta/\delta_{\Gamma}|$. The plot 
at the left has to be compared with the prediction of Eq. (\ref{zetafin4}) (appropriate for  a transition dust--radiation transition), while the plot at the right has to 
be compared with 
Eq. (\ref{zetafin5}) (appropriate for a stiff--radiation transition). In both 
plots the predictions of, respectively, Eqs. (\ref{zetafin4}) and (\ref{zetafin5}) 
is illustrated with a dashed line. For comparison, we also report 
the expectation obtainable in the limit $\epsilon \to 0$.  The 
numerical correctness of Eq. (\ref{zetafinal}) seems then numerically justified.
In Fig. \ref{F5} the integration of the evolution equations of the fluctuations 
is reported for different values of $\epsilon$ and $\delta_{\Gamma}$.

In more general terms one can think of cases when 
$\xi_{\rm a}=\epsilon \rho_{\rm a}^{m}$. In the absence of a decay 
rate ( and for a single fluid) \cite{barrow1}, if $m > 1/2$, there exist general deflationary solutions 
beginning in a quasi-de Sitter stage that evolve away from it for large 
times. Always in the absence of decay rate (and always in the single fluid case) \cite{barrow1}, if $m < 1/2$  the solution 
starts expanding and reaches, for large times, a quasi-de Sitter solution. These 
statements hold, as stressed, for the case of a single fluid and in the 
absence of a decay rate. If there are many fluids, allowed to decay, the 
evolution depends  upon the relative balance of the different 
parameters of the model. The main end of the present approach is to investigate the 
situations when the bulk viscosity coefficient does not dominate for large 
positive times.  In this sense, solutions that do not lead to a 
radiation-dominated background for $t > 1/\overline{\Gamma}$ are less interesting 
for the present analysis (but may be relevant in related contexts).  

\begin{figure}
\begin{center}
\begin{tabular}{|c|c|}
      \hline
      \hbox{\epsfxsize = 7 cm  \epsffile{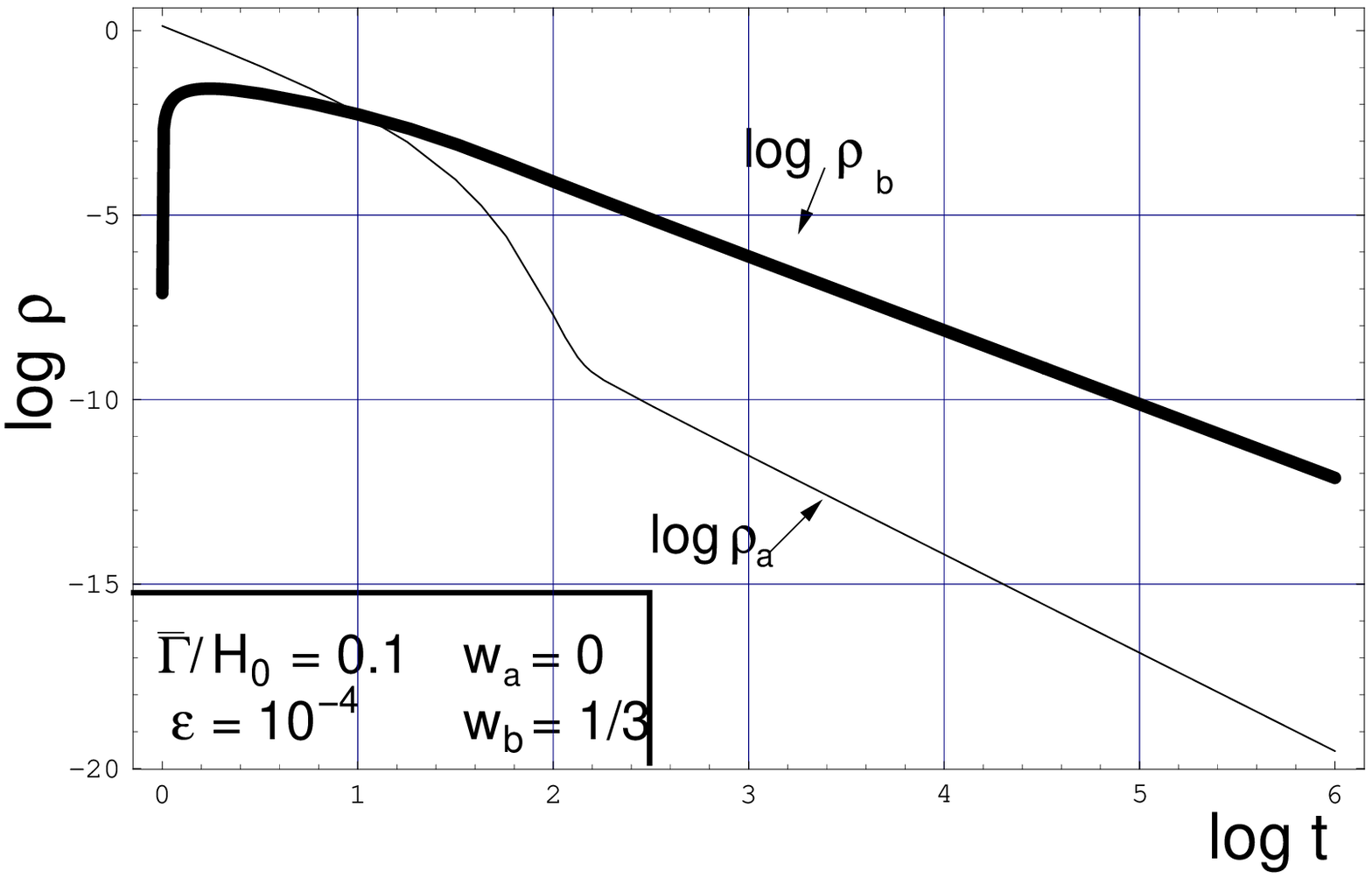}} &
      \hbox{\epsfxsize = 7 cm  \epsffile{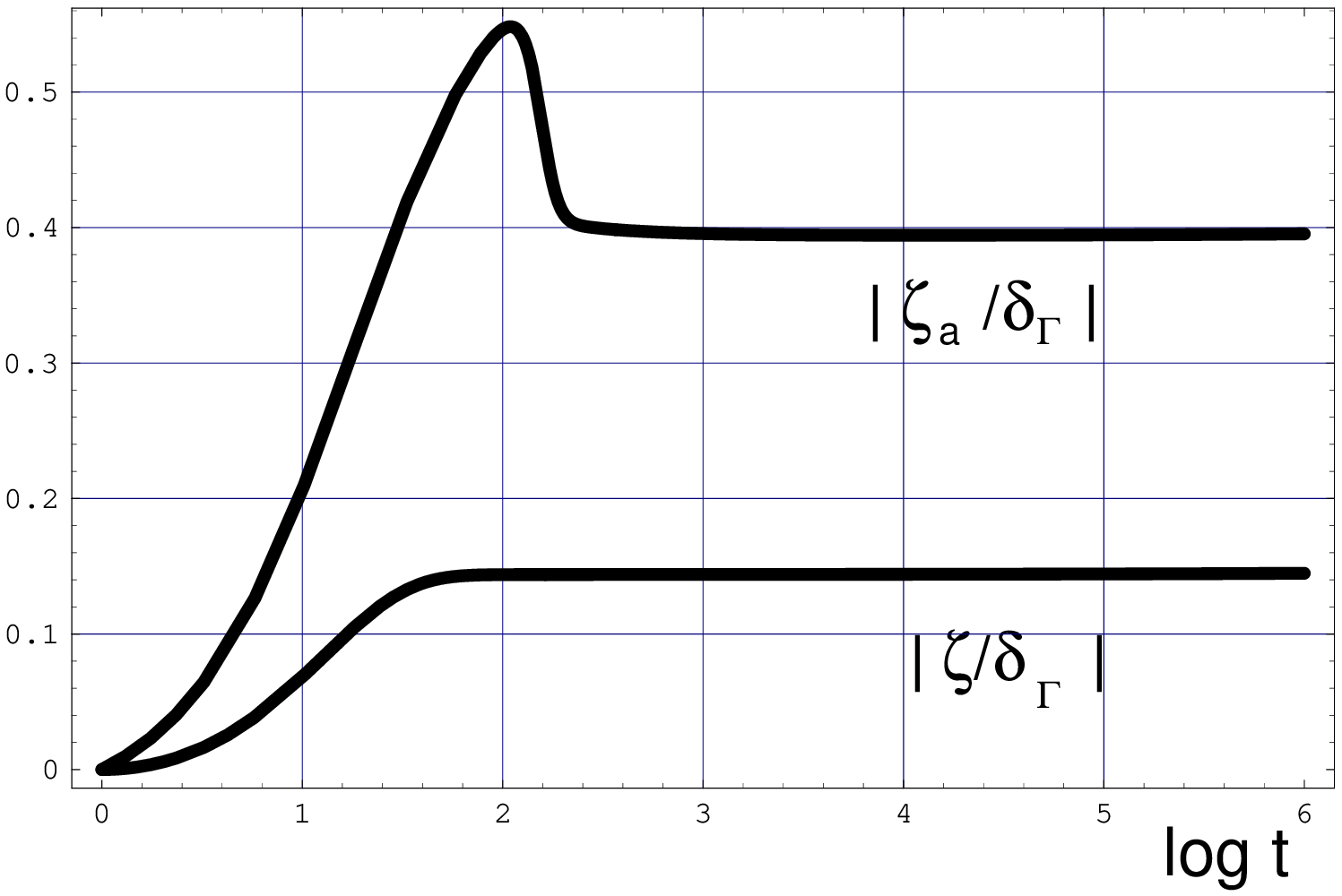}}\\
      \hline
\end{tabular}
\end{center}
\caption[a]{The integration of the system is illustrated in the case $\xi_{\rm a} =
\epsilon \rho_{\rm a}^{1/4}$ and $\xi_{\rm b} =0$.  The value $\delta_{\Gamma} = 10^{-8} $ has been used. In the  plot on the 
left-hand side the evolution of the background energy density 
is reported on a double logarithmic scale. On the right-hand-side plot the evolution of $\zeta$ and $\zeta_{\rm a}$ is reported for the same background values used  in the left hand one.}
\label{F6}
\end{figure}

In this perspective, even if $m < 1/2$,  it could happen that radiation dominates 
very quickly if the decay rate is sufficiently large and $\epsilon$ is sufficiently 
small. Consider, for instance, the case illustrated in Fig. \ref{F6},
where we took $\xi_{\rm a} = \epsilon \rho_{\rm a}^{1/4}$.
If $\overline{\Gamma}/H_{0} = 0.1 $ and $\epsilon= 10^{-3}$ 
the evolution equations can be integrated. To be safe, the evolution equations 
of the fluctuations may be integrated directly in the off-diagonal gauge 
in terms of $\phi$ (not of $\zeta$). This choice has the advantage 
that the evolution of $\phi$ is never singular even if $\dot{H}\to 0$ 
during a possible quasi-de Sitter stage. 
Also other cases of this rich system may be physically relevant. Those
presented here were selected because they illustrate, in simple terms, the interplay 
between bulk viscous stresses and fluctuating decay rates.

\renewcommand{\theequation}{7.\arabic{equation}}
\section{Concluding remarks}
\setcounter{equation}{0}
Bulk viscous effects of interacting relativistic  fluids 
have been investigated. These two effects were never considered 
simultaneously, to the best of our knowledge. 
The energy-momentum tensors of each of the 
fluids of the relativistic plasma have been parametrized as the sum 
of an inviscid contribution and of a bulk  viscous correction.   
Different results have been obtained:
\begin{itemize}
\item{}  a fully gauge-invariant description of the evolution of the fluctuations 
of the geometry has been discussed in the presence of both a 
fluctuating decay rate and  a fluctuating bulk viscosity coefficient;
\item{} gauge-dependent descriptions have been derived and exploited with 
particular attention to the off-diagonal (or uniform-curvature) gauge.
\item{} if the bulk viscosity coefficient and the decay rate 
are allowed to have spatial variation on super-horizon scales, 
curvature fluctuations may be generated even if they are initially 
vanishing;
\item{} examples of the latter phenomenon  were 
provided both analytically and numerically.
\end{itemize}
As far as applications are concerned, two different situations 
may arise. In the first case the bulk viscosity is absent from the 
background, but its inhomogeneities may affect the evolution of 
the fluctuations with wavelengths larger than the Hubble radius.
In the second case, the bulk viscosity does contribute 
to the background. It has been shown that situations can be envisaged 
where the contribution of the bulk viscosity coefficient becomes 
subleading after the decay of the viscous fluid of the mixture. In this case, 
the curvature fluctuations for length scales larger that the Hubble radius 
are determined by the interplay of the fluctuating viscosity and of the 
inhomogeneous decay rate. The final asymptotic value of the curvature 
fluctuations has been computed in specific numerical examples, where, 
for instance, the decay products are in the form of a radiation fluid. 
\newpage

\begin{appendix}
\renewcommand{\theequation}{A.\arabic{equation}}
\setcounter{equation}{0}
\section*{Appendix: vector modes}
For completeness, 
some results concerning the evolution 
of vector modes in a mixture of viscous interacting 
fluids are reported in the present appendix. They are not central 
to the present analysis, but  are logically 
connected to the considerations presented so far. 
In fact, one may wonder if effects similar to the ones 
discussed for the scalar modes may also arise 
in the case of vector modes. 
Consider then the vector fluctuations of the four-dimensional metric (\ref{bm}).
They are parametrized by two (solenoidal) three-dimensional 
vectors $Q_{i}$ and $W_{i}$ which have, overall, three 
independent components:
\begin{eqnarray}
&& \delta_{\rm v} g_{0 i} = - a^2 Q_{i},\,\,\,\,\,\,\,
\delta_{\rm v} g^{0 i} = - \frac{ Q^{i}}{a^2},
\nonumber\\
&& \delta_{\rm v} g_{i j} = 
a^2 ( \partial_{i} W_{j} + \partial_{j} W_{i}),\,\,\,\,\,
 \delta_{\rm v} g^{i j} = -
\frac{1}{a^2}( \partial^{i} W^{j} + \partial^{j} W^{i}),
\label{VF}
\end{eqnarray}
where $\delta_{\rm v}$ denotes the vector fluctuation 
of the various entries of the perturbed metric. It is possible 
to select a gauge where $\delta_{\rm v} g_{ij}$ is vanishing. 
This choice amounts to fixing two degrees of freedom 
of the perturbed metric.  The vector fluctuations of the Einstein equations 
as well as the vector fluctuations of the equations describing the energy-momentum exchange between different fluids can be written as 
\begin{eqnarray}
&& \delta_{\rm v} {\cal G}_{\mu}^{\nu} = 8\pi G \delta_{\rm v} {\cal T}_{\mu}^{\nu},
\label{VP1}\\
&& \nabla_{\mu}\delta_{\rm v} {\cal T}^{\mu\nu}_{\rm a} = - \overline{\Gamma} 
\delta_{\rm v} g^{\nu \alpha} \overline{u}_{\alpha} ( p_{\rm a} + \rho_{\rm a})
- \overline{\Gamma} 
\overline{g}^{\nu \alpha} \delta_{\rm v} u_{\alpha} ( p_{\rm a} + \rho_{\rm a}),
\label{VP2}\\
&& \nabla_{\mu}\delta_{\rm v} {\cal T}^{\mu\nu}_{\rm b} = \overline{\Gamma} 
\delta_{\rm v} g^{\nu \alpha} \overline{u}_{\alpha} ( p_{\rm a} + \rho_{\rm a}) +
 \overline{\Gamma} 
\overline{g}^{\nu \alpha} \delta_{\rm v} u_{\alpha} ( p_{\rm a} + \rho_{\rm a}),
\label{VP3}
\end{eqnarray}
with the same notation as used in Eqs. (\ref{BV1}), (\ref{BV2}) and (\ref{sum}),
but with the difference that, here, vector fluctuations are considered. 

Using the perturbed metric (\ref{VF}) and recalling that $\overline{u}_{0}\delta_{\rm v} u^{i} = {\cal V}^{i}$,  
the $(0i)$ and $(ij)$ components of Eq. (\ref{VP1}) imply:
\begin{eqnarray}
&& \nabla^2 Q_{i} = - 16\pi G ( \rho + {\cal P}) a^2 {\cal V}_{i},
\label{VP4}\\
&& Q_{i}' + 2 {\cal H} Q_{i} =0. 
\label{VP5}
\end{eqnarray}
Equations (\ref{VP2}) and (\ref{VP3}) lead, instead, to the following pair 
of equations:
\begin{eqnarray}
&& \bigl[ {\cal V}_{ i}^{({\rm a})} (\rho_{\rm a} + {\cal P}_{\rm a})\bigr]' 
+ 4 {\cal H} \bigl[ {\cal V}_{i}^{({\rm a})} ( \rho_{\rm a} + {\cal P}_{\rm a} ) \bigr]
= - a \overline{\Gamma} (\rho_{\rm a} + p_{\rm a}) {\cal V}_{i},
\label{VP6}\\
&& \bigl[ {\cal V}_{ i}^{({\rm b})} (\rho_{\rm b} + {\cal P}_{\rm b})\bigr]' 
+ 4 {\cal H} \bigl[ {\cal V}_{i}^{({\rm b})} ( \rho_{\rm b} + {\cal P}_{\rm b} ) \bigr]
=  a \overline{\Gamma} (\rho_{\rm a} + p_{\rm a}) {\cal V}_{i}.
\label{VP7}
\end{eqnarray}
Considerations similar to the ones reported in Eqs. (\ref{TOTth}), (\ref{TOTxi1}) and (\ref{TOTxi3}) lead to the evolution equation 
of the total velocity field ${\cal V}_{i}$, i.e.
\begin{eqnarray}
&&  \bigl[ {\cal V}_{ i} (\rho_{\rm a} + {\cal P}_{\rm a})\bigr]' 
+ 4 {\cal H} \bigl[ {\cal V}_{i} ( \rho_{\rm a} + {\cal P}_{\rm a} ) \bigr]=0,
\nonumber\\
&& (\rho +{\cal P}) {\cal V}_{i} = (\rho_{\rm a} + {\cal P}_{\rm a}) {\cal V}_{i}^{(\rm a)} + (\rho_{\rm b} + {\cal P}_{\rm b}) {\cal V}_{i}^{(\rm b)}. 
\label{consV}
\end{eqnarray}

From Eqs. (\ref{VP4})--(\ref{VP7}) it can be easily argued that the 
spatial fluctuations of the bulk viscosity coefficient do not 
contribute, as expected, to the evolution of the vector 
fluctuations of the geometry;  the situation is then 
very similar to the one discussed in the inviscid case in a number 
of studies that appeared  in the past \cite{B1a,B2a} as well as more recently \cite{battbrand,vmax1,vmax2,batteas}. The difference is, of course, that 
in the inviscid case the evolution equations of the background are 
different from those where a homogeneous bulk viscosity 
coeffcient is included. Finally, always for completeness, we should mention 
that the tensor modes, in some classes of bulk viscous solution, have 
been analysed in \cite{maxtens}.
\end{appendix}

\end{document}